# The impact of hydrostatic pressure, nonstoichiometry, and doping on trimeron lattice excitations in magnetite during axis switching

T. Kołodziej[1,2], J. Piętosa[3], R. Puźniak[3], A. Wiśniewski[3], G. Król[1,4], Z. Kąkol[1], I. Biało[1,5], Z. Tarnawski[1], M. Ślęzak[1], K. Podgórska[1], J. Niewolski[1], M. A. Gala[1,6], A. Kozłowski[1], J. M. Honig[7], W. Tabiś[1,6*]

[1]*AGH University of Krakow, Faculty of Physics and Applied Computer Science, Aleja Mickiewicza 30, 30-059 Kraków, Poland*
[2]*SOLARIS National Synchrotron Radiation Centre, Czerwone Maki 98, 30-392 Kraków, Poland*
[3]*Institute of Physics, Polish Academy of Sciences, Aleja Lotników 32/46, 02-668 Warszawa, Poland*
[4]*AGH University of Krakow, IT Solutions Centre, Aleja Mickiewicza 30, 30-059 Kraków, Poland*
[5]*Physik-Institut, Universität Zürich, Winterthurerstrasse 190, CH-8057 Zurich, Switzerland*
[6]*Institute of Solid State Physics, TU Wien, 1040 Vienna, Austria*
[7]*Department of Chemistry, Purdue University, West Lafayette, IN, USA*
[*] Email: wtabis@agh.edu.pl

**Abstract**
Trimeron lattice excitations in single crystalline magnetite, in the form of the *c* axis switching (i.e. the reorganization of the lattice caused by external magnetic field) at temperatures below the Verwey temperature $T_V$ are observed by magnetization experiments. These excitations exhibit strong sensitivity to doping (with Zn, Al, and Ti), nonstoichiometry and hydrostatic pressure ($p < 1.2$ GPa). The considered indicators of the axis switching (AS) are: the switching field $B_{sw}$, the energy density needed to switch the axis $E_{sw}$ and the activation energy $U$. Our results show that hydrostatic pressure $p$ weakens the low-$T$ magnetite structure (decreases $T_V$) and has roughly similar effects on AS in Zn-doped $Fe_3O_4$ and, in much less extent, in stoichiometric magnetite. We have, however, found that while doping/nonstoichiometry also lowers $T_V$, making it more prone to temperature chaos, it drastically increases the switching field, activation and switching energies suggesting that the trimeron order, subject to change while AS occurs, is more robust. Consequently, we conclude that the manipulation of trimerons in the process of axis switching and the mechanisms leading to the Verwey transition are distinct phenomena.

## 1. Introduction

The charge and orbital order in magnetite at $T < T_V$, where $T_V$ is the Verwey temperature, exhibits characteristic cigarlike structures in a form of polarons known as trimerons [1]. Recently, spectroscopic signatures of the low-energy electronic excitations of the trimeron network were reported [2]. In seemingly unrelated experiments, we have been observing the magnetic field-induced manipulation of the monoclinic *c* axis in the process referred to as "axis switching" (AS). Since AS also impacts the trimeron order, this phenomenon represents an alternative method to excite the trimerons. In this study, we present the investigation of these excitations under various conditions, including sample defect structure (nonstoichiometry, Zn, Al, and Ti doping) and hydrostatic pressure up to 1.2 GPa. These investigations were performed using magnetic moment vs. applied magnetic field, $m(B)$, measurements.

Magnetic axis switching in magnetite was first mentioned and analysed by Calhoun in 1954 [3] in his magnetization vs. magnetic field experiments, but the phenomenon was also observed in other materials [4,5,6]. Below the isotropy point of $T_{IP} = 130$ K (where anisotropy energy in magnetite is minimal) the easy magnetization axis is in one of <100> directions (see e.g. [7]) and this is preserved when cooling magnetite below $T_V$ at which point cubic symmetry turns to monoclinic: each cubic <100> may become the monoclinic *c* axis and, simultaneously, the easy magnetic axis. Therefore, the material breaks into structural domains on cooling [three of them, with different *c* axes, are shown symbolically in Fig. 1a[1]] [8,9]. An external magnetic field $B > 0.2$ T applied along one of the <001> crystallographic axes during cooling across $T_V$ may force the magnetic easy axis along this particular direction [8] [Fig. 1b], what is proved by the characteristic $m(B)$ relation [Fig. 1c]. If the sample is now magnetized along the other <100> direction below $T_V$ (and at $T$ higher than 50 K), a reorientation of magnetic moments, i.e., axis switching, takes place and this <100> direction becomes a new easy axis [3,8,9,10].

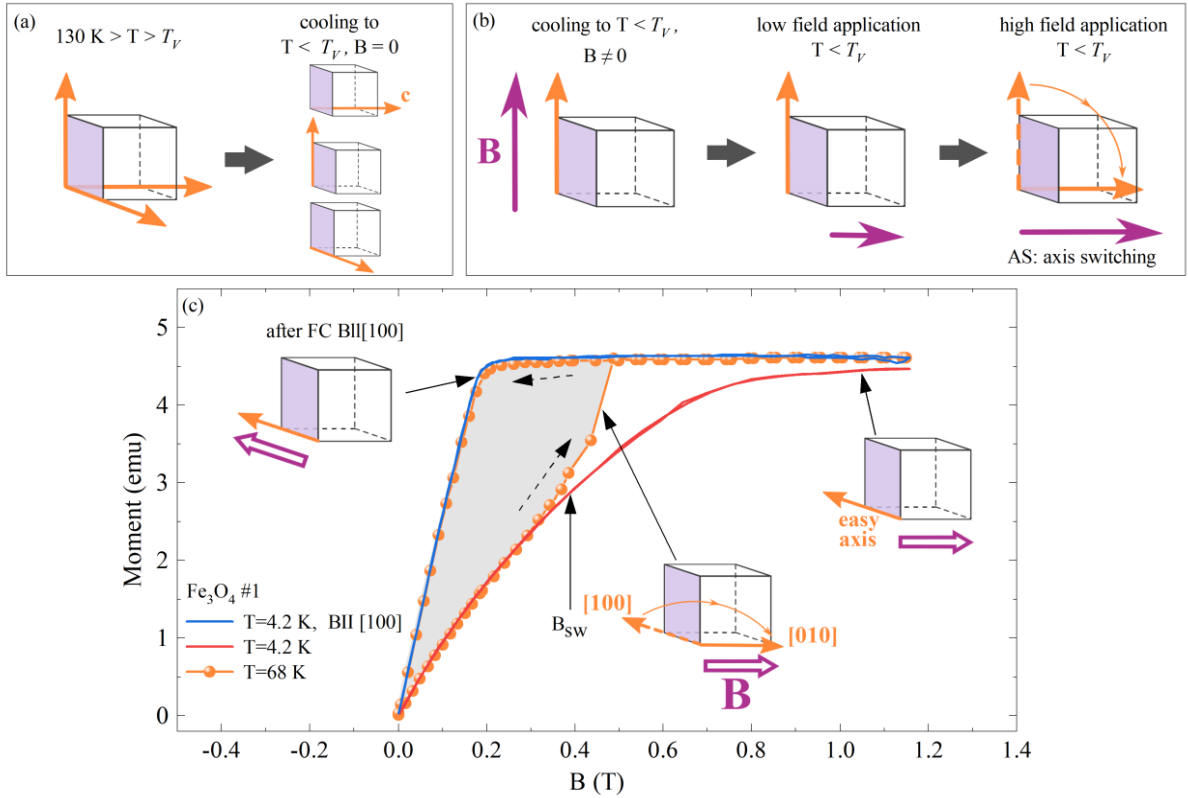

FIG. 1. Cooling a magnetite sample below $T_V$ results either with three principle domains of monoclinic $Cc$ structure with $c$ easy axes along previous cube edges (a), or one $c$ axis (b) when the external magnetic field along one of the cube edges is applied on cooling. Subsequent field application (at $T > 50$ K) along one of the two other cube directions initially has no effect, but eventually can switch irreversibly the axis that can best be observed in $m$ vs $B$ experiment (c); here the dashed arrows indicate the change of magnetic field, while the bold arrow points to the switching field $B_{sw}$. The shaded area in the plot in panel (c) is a measure of energy needed to switch the axis, i.e. is proportional to the switching energy density $E_{sw}$. The figure is a part of $m$ vs $B$ measurement for sample #1 (the complete data are presented in Fig. 3Sa in Supplemental Material (SM) [11].

Calhoun showed [3] that the magnetic field $B_{sw}$ needed to switch the axis depends on temperature $T$, activation energy $U$, and is observed when the process starts to have a collective behavior (this is depicted by the $C$ parameter; only above some threshold $C$ value, the process starts to be experimentally observed). It all resulted in the characteristic formula $B_{i\_sw} = CT\exp(U/k_B T)$ ($B_{i\_sw}$ in the internal switching field) that links all the parameters mentioned above.

Subsequent studies have revealed that not only the magnetic axes but also the crystal axes, and thus the trimeron lattice, undergo irreversible changes in response to the external magnetic field. We have extensively explored this phenomenon using various experimental techniques, including global probes such as magnetization [8], symmetry oriented techniques such as x-ray diffraction (XRD) [9] and resonant x-ray scattering (RXS) [12], as well as microscopic methods such as nuclear magnetic resonance (NMR) [13] and Mössbauer spectroscopy (MS) [12]. In essence, these experiments demonstrated that each manipulation of the $c$ axis is an excitation of the trimeron lattice.

---

[1] Note that the symbolic picture Fig.1 shows only main monoclinic domains, with the $c$ axes roughly along one of the cube axes, a much more structurally complicated picture emerges with 24 structural domains present. Although these domains are not immediately observed in $m(B)$ experiments, they certainly add to the behavior of trimeron reconfiguration.

It is worth noting that the trimeron lattice is not only a structural building block but also a fundamental element of magnetite's electronic structure. As recently demonstrated, the trimeron lattice critically softens at $T_V$, indicating its intimate link with the Verwey transition [2]. Furthermore, the transition is highly sensitive to even minor perturbations of the lattice, such as doping and nonstoichiometry. For instance, already 1/3% of Fe ions are already either vacant or replaced with Zn, and Ti lowers $T_V$ by 15 K. Higher doping changes the character of the transition (refer to Sec. I in [11]), and 1% suppresses the transition entirely. Hydrostatic pressure also exerts a significant influence on the transition by reducing $T_V$.

Hence, the observation of AS, which represents an alternative means of modifying the trimeron system, and investigating how this process depends on sample doping and pressure can provide valuable insights into the understanding of the switching mechanism itself. Moreover, these investigations can contribute to a better understanding of the trimeron structure, trimeron excitation spectrum and, ultimately, the Verwey transition (VT). It can, also, help to describe possible trimeron order that exists, in some extent, above $T_V$, where it is dynamic rather than static [14], or its potential role in creating electronic nematic order characterized by static, low-value electronic disproportionation, as recently proposed [15].

This study primarily focuses on the excitations of the trimeron lattice and the mechanism underlying the AS. We explore how AS, represented by changes in the $m(B)$ behavior, responds to various kinds of defects (nonstoichiometry, Zn, Al, and Ti doping), or hydrostatic pressure. We have deliberately selected nonmagnetic dopants to avoid introducing additional variables; however, these dopants possess distinct valences and locate at different lattice positions. Finally, the defect concentration (dopants and Fe vacancies) was high enough to considerably alter $T_V$ while maintaining the discontinuous character of the VT. Low hydrostatic pressure also affects $T_V$ by inducing changes in the electronic, rather than the defect structure, thus providing insights into the electronic aspects of AS. Additionally, we have investigated the response of the trimeron order to the sequential manipulation of the sample structure, referred to as "sample training"; this is shown in [11].

We investigated several key parameters that provide quantitative characterization of the AS process. These parameters include the switching field $B_{sw}$ and switching energy density $E_{sw}$ [see Fig. 1c], both directly derived from the $m$ vs $B$ experiments, as well as the fitting parameter $U$, which represents the activation energy. We have discovered that through increasing the temperature, increasing number of structural twins (caused by sample training, see Sec. 4(d) in [11]), or by applying hydrostatic pressure, the trimeron lattice becomes more susceptible to changes, reflected in decreasing the switching field $B_{sw}$ and switching energy $E_{sw}$. On the contrary, doping or nonstoichiometry appear to strengthen the trimeron order increasing $B_{sw}$ and $E_{sw}$. Furthermore, the activation energy $U$ exhibits an increase with doping or nonstoichiometry, while the impact of pressure on $U$ differs between doped and stoichiometric magnetite. Specifically, $U$ appears to be pressure independent for stoichiometric magnetite, whereas it decreases under pressure for Zn-doped sample.

The structure of the paper is as follows: In Sec. 2, we present the characterization of the sample and the experimental procedures. This is followed by the presentation of results of AS in different scenarios, including stoichiometric and nonstoichiometric magnetite as well as Zn-, Al-, and Ti-doped samples and, finally studies conducted under hydrostatic pressure (Sec. 3). Our findings are discussed in Sec. 4 and the article is concluded in Sec. 5.

## 2. Samples and experimental procedure
Single crystals of stoichiometric, nonstoichiometric, and doped magnetite were grown from the melt by the cold crucible technique (skull melter) at Purdue University, USA [16]. The crystals were subsequently annealed to achieve the appropriate metal to oxygen ratio [17,18]. The samples included magnetite with varying levels of nonstoichiometry, $3\delta$ in $Fe_{3(1-\delta)}O_4$ with values of $3\delta = 0$, 0.0075 and 0.0105. In addition, doping was performed with zinc, $Fe_{3-x}Zn_xO_4$, ($x = 0.0066$, $x = 0.007$ and $x = 0.01$), aluminum, $Fe_{3-x}Al_xO_4$, ($x = 0.012$), and titanium $Fe_{3-x}Ti_xO_4$, ($x = 0.01$). It is important to note that the chosen doping levels and nonstoichiometry were deliberately kept small to only subtly alter the properties of magnetite and maintain a discontinuous character of the Verwey transition ($x = 3\delta \leq 0.012$). The effects of doping and nonstoichiometry on A (tetrahedral) and B (octahedral) sites in magnetite structure, as well as their impact on the Verwey transition in general, are elaborated in Sec. 1 of [11] and Refs. [18,19,20,21,22,23] therein. To assess the quality of the samples, the width of the Verwey transition in the temperature dependence of ac susceptibility was examined. The results of these tests, along with the respective values of $T_V$, along with selected details of the samples, are presented in the Table 1, and furthermore illustrated graphically in Fig. 1S of [11].

All measurements described below were performed using Vibrating Sample Magnetometer [(VSM); Princeton Applied Research, PAR Model 4500 with cryostat Model 153 and Varian 12-in. electromagnet, controlled by the Lakeshore VSM Controller 7300]. In the case of the experiments conducted under pressure (up to 1.2 GPa), the description of the cell and the setup can be found in Ref. [24], with some details presented in [11] (Fig. 3S).

TABLE I. Results of the samples characterization ($T_V$; Zn, Al, Ti concentration and $\delta$) and fitted parameter $U$ describing temperature dependence of the switching field. Zn and Ti concentration $x$ were drawn from $T_V$ vs $x$ relation (Fig. 1S(b) in SM), Al content was measured by x-ray microprobe, and $\delta$ was set by the annealing conditions.

| Sample | $T_V$ (K) | $U/k_B$ (K) | Uncertainty of U $\Delta(U/k_B)$ (K) |
|---|---|---|---|
| $Fe_3O_4$ #1 (cylinder) | 123.7 | 358 | 14 |
| $Fe_3O_4$ #2 (sphere) | 123.8 | 470 | 23 |
| $Fe_3O_4$ #3 (cylinder) used for pressure studies | 124 | 337 | 11 |
| $Fe_{3(1-\delta)}O_4$, $3\delta = 0.0075$ | 114.8 | 747 | 60 |
| $Fe_{3(1-\delta)}O_4$, $3\delta = 0.0105$ | 109.7 | 846 | 35 |
| $Fe_{3-x}Zn_xO_4$. Zn #1 ($x_{Zn} = 0.007$) | 114.4 | 584 | 10 |
| $Fe_{3-x}Zn_xO_4$. Zn #2 ($x_{Zn} = 0.01$) | 110.8 | 688 | 54 |
| $Fe_{3-x}Zn_xO_4$. Zn #3($x_{Zn} = 0.0066$) used for pressure studies) | 114.5 | 490 | 20 |
| $Fe_{3-x}Al_xO_4$, $x_{Al} = 0.012$ | 112.8 | 682 | 50 |
| $Fe_{3-x}Ti_xO_4$, $x_{Ti} = 0.01$ | 111.4 | 900 | 200 |

In all magnetization experiments (one sphere and nine cylinders of various radius), the cylinder axis (and one of the <100> axes in the case of a sphere) was set along the VSM vertical probe (note that cubic notation will be used throughout the rest of the paper). Two other <100> directions in the horizontal plane were identified at 290 K in 0.15-0.3 T field as these directions with the smallest moment (since these directions were hard axes in the high $T$ phase of magnetite), see Fig. 4S(b) and 4S(f) in SM [11]. After horizontal <100> axes were found, the samples were cooled to the specified temperature below $T_V$ in magnetic field larger than 0.5 T (which was proved to be strong enough to define the $c$ axis [8]; see Sec. 2 and Fig. 2S in SM [11] for further discussion) set along the [100] direction (as schematically shown in Fig. 1). This field-cooling (FC) procedure establishes an easy axis in this particular direction, what is confirmed by the $m(B)$ measurement with $B$ along this direction [Fig. 1c, blue line].

The subsequent steps were as follows:
* The $m$ vs. $B$ measurements were typically conducted at a specified temperature with a magnetic field applied along the initially defined easy axis [this is the blue line in Fig. 1(c)]. From the initial part of the $m(B)$ curve, the demagnetization factor $D$ was calculated, which was used later to determine the internal field $B_{i\_sw}$; in this linear region of the $m(B)$ curve the external field compensates the demagnetizing field.

* The sample was then rotated by 90° to align the magnetic field along the other <100> direction, referred to as [010], an unspecified magnetic direction below $T_V$. The resulting $m(B)$ curve obtained at 4.2 K is presented as the red line in Fig. 1(c).
* At sufficiently high temperature, the axis switching occurred, eventually setting the new easy axis direction along the applied magnetic field, represented by orange dots in Fig. 1(c).
* The sample was subsequently heated above $T_V$ and again field cooled along the [100] direction to a new temperature below $T_V$. At this temperature, measurements of $m(B)$ along [100] and along [010] were repeated.

## 3. Experimental results

### a. Ambient pressure data

The switching field $B_{sw}$, which is the most direct experimental parameter extracted from our measurements, was determined from each of the $m(B)$ curves acquired along the magnetically unspecified axes. The recalculation of $B_{sw}$ into the intrinsic field $B_{i\_sw}$ was performed using the demagnetization parameters $D$, see Fig. 1. $B_{i\_sw}$ vs $T$ plots for all the studied samples, including those under elevated pressure, are shown in Fig. 2. Although cumulative figures, like this one, best present the clear tendency of the parameters drawn from the measurements performed on all of the samples, for clarity, $B_{i\_sw}$ vs $T$ is further plotted separately for samples under ambient pressure, and the pressure dependence of $B_{i\_sw}$ in Fig. 7S(a) and Fig. 12S(a) of [11], respectively.

All the data were extracted from the results of magnetic moment vs magnetic field measurements performed at several temperatures for three stoichiometric, two nonstoichiometric, as well as in three Zn-doped, one Al-doped, and one Ti-doped samples, under ambient and increased pressure; the results are presented in Figs. 4S, 5S, 6S, 9S and 11S in [11] and in Fig. 5 below.

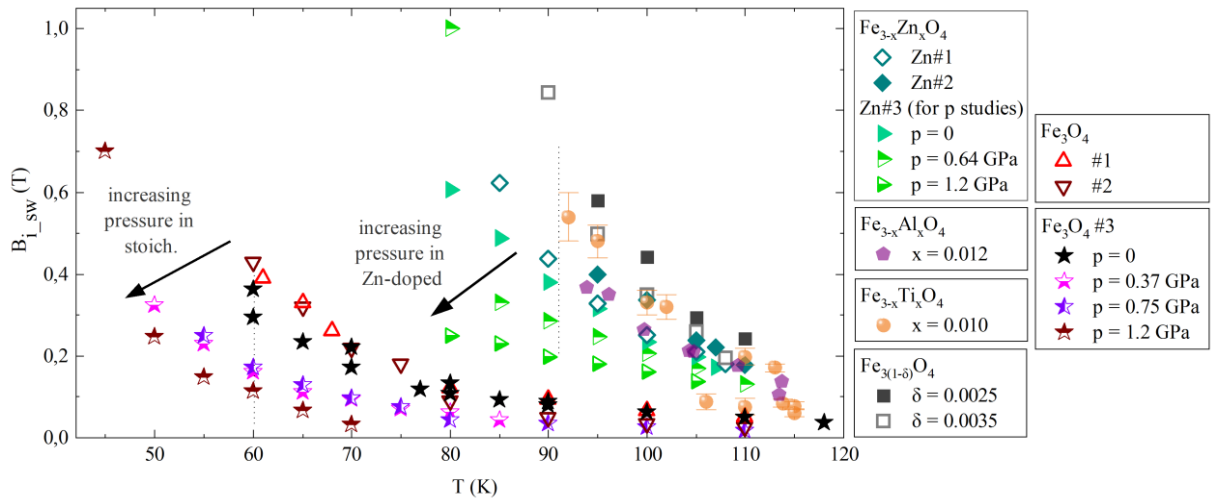

FIG. 2. Temperature dependence of the switching field, $B_{i\_sw}$ for all studied samples. The data for separate cases: for samples under ambient pressure and pressure dependence for #3 and Zn #3 are also presented in [11] [Figs. 7S(a) and 12S(a), respectively].

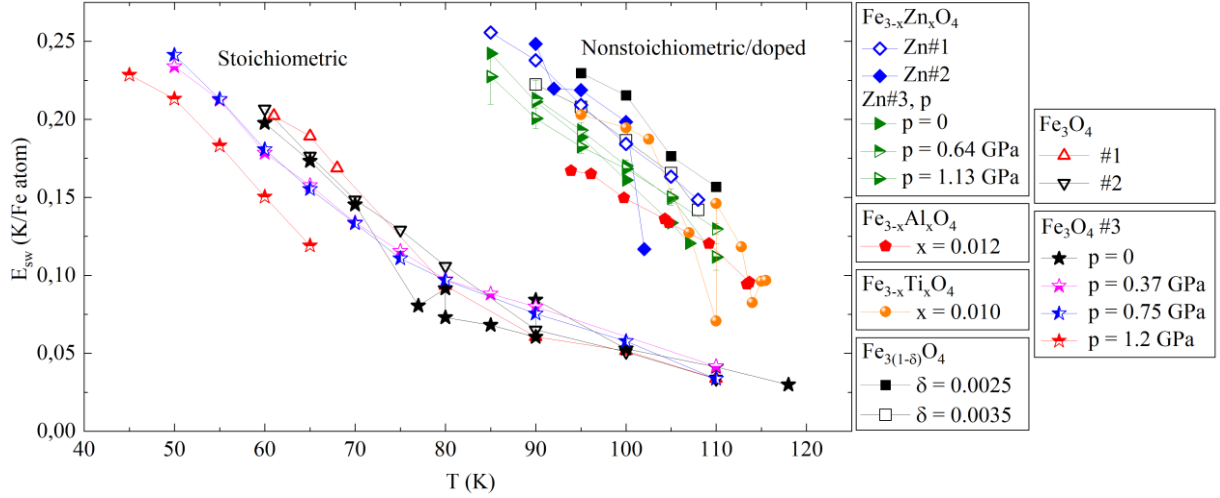

FIG. 3. Temperature dependence of the energy density $E_{sw}$ (in Kelvin per one Fe atom) required to switch the axis. Representative error bars are shown for two cases (sample Zn #3, $p = 0.64$ and $1.13$ GPa). The data for separate cases: for samples under ambient pressure and pressure dependence for #3 and Zn #3 are given in Figs. 7S(b) and 12S(b), respectively.

In Fig. 3, we illustrate the energy (work), $E_{sw}$, needed to switch the easy $c$ axis, estimated from the $m(B)$ curves [see Fig. 1(c) for $E_{sw}$ definition]. The temperature dependence of $B_{i\_sw}$, represented as $\ln(B_{i\_sw}/T)$ vs $1000/T$, was fitted with Calhoun's formula $B_{i\_sw} = CT\exp(U/k_B T)$. The activation energy $U$, one of the fitting parameters which is the second important value characterizing AS, is shown in Fig. 4 as a function of $T_V$. The fits are presented in Figs. 4S, 5S, 6S, 9S, 11S in [11] and in Fig. 5.

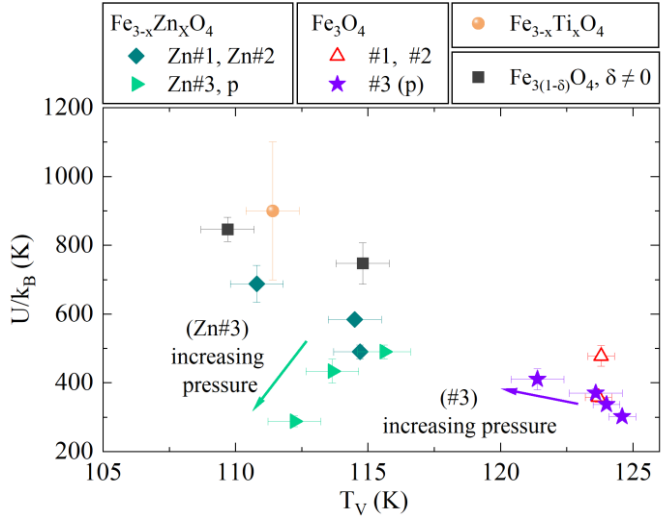

FIG. 4. Correlation between activation energy $U$ and the Verwey transition temperature $T_V$.

### b. Experiments at elevated pressure

Two samples, one stoichiometric magnetite, $Fe_3O_4$ #3, and one Zn doped ($Fe_{3-x}Zn_xO_4$ Zn #3, see Fig. 6S(c), and 6S(h) in [11], where the ambient pressure results for this sample are shown), were measured under elevated pressure. After the <100> directions perpendicular to the cylinder axis were found (which was done at 160 K to let the transmitting pressure oil to freeze), a sample was field-cooled (FC) down to the lowest temperature (below 20 K). Representative results of $m(B)$ for stoichiometric sample #3 and under $p = 0.75$ GPa are shown in Fig. 5(a) (complete results are provided in Figs. 9S and 11S in [11]).

As before, $U$ was estimated from $\ln(B_{i\_sw}/T)$ vs. $1000/T$ curves, shown in Fig. 5(b) (stoichiometric sample) and in Fig. 5(c) (Zn doped).

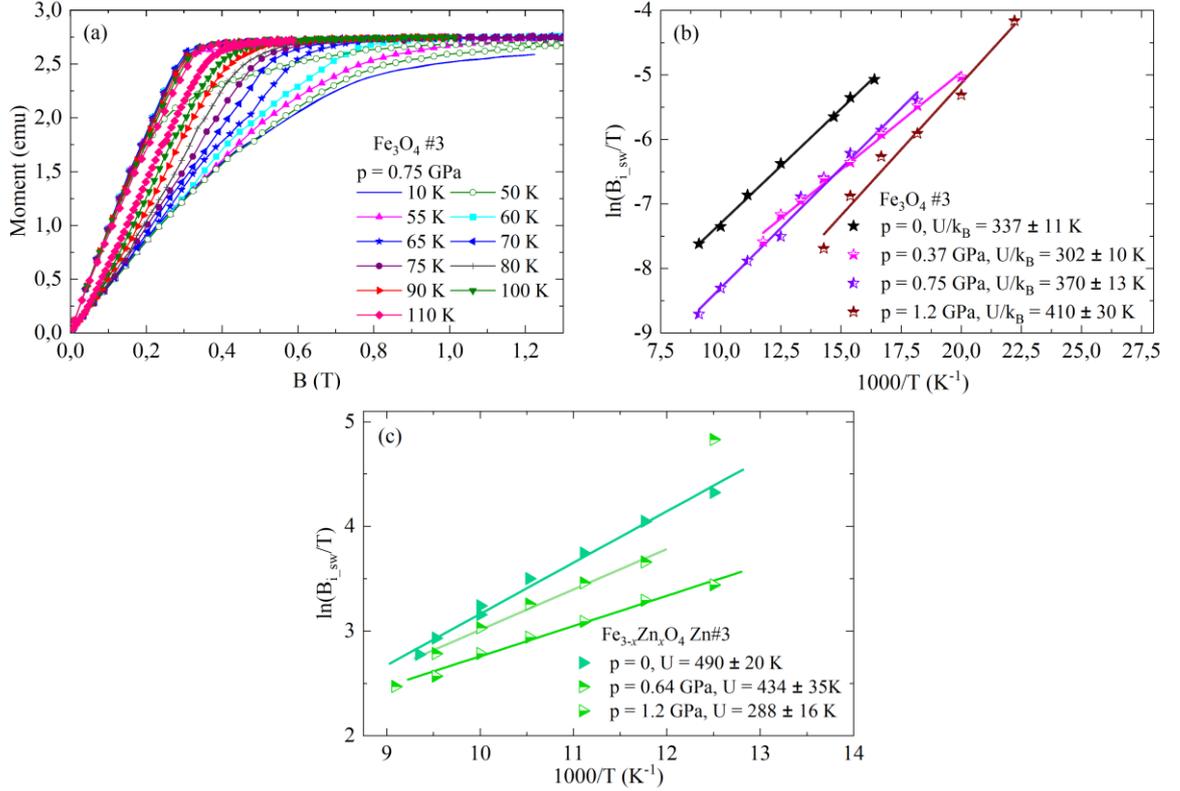

FIG. 5. Representative results of pressure (a) measurements of the stoichiometric sample $Fe_3O_4$ #3 ($m(B)$ for ambient pressure is in Fig. 4S(d) in [11]).). In panels (b), (c) $\ln(B_{i\_sw}/T)$ vs $1000/T$ and the ensuing $U$ values for all pressure values are presented for $Fe_3O_4$ #3 and Zn-doped magnetite, respectively. See SM [11] for the collection of all data and more discussions of the results for 1.2 GPa in $Fe_3O_4$ #3.

The procedure of setting the easy direction along the cooling field proved to be effective in the case of the pressure $p = 1.2$ GPa for undoped sample $x = 0$ at temperatures below 65 K. However, at higher temperatures, above ca. 70 K, $m(B)$ along [100] (expected to be an easy axis after FC) did not exhibit typical behavior of an easy axis. This phenomenon will be further discussed in Sec. 4 c below.

Additionally, the results pertaining to the pressure dependence of the magnetic moment step at $T_V$ (aimed to determine the Verwey transition temperature vs $p$) and pressure dependence of magnetic moment are presented in Figs. 13S and 14S in [11], where our data are further discussed in the context of the results published in Refs. [25,26,27,28,29,30,31] therein.

## 4. Discussion

An effort was already made to observe AS more microscopically, by NMR [13] or Mössbauer spectroscopy [12], to see the individual Fe positions that either change during AS, as $B$ positions, or observe the ongoing process without changing their valence (as $A$ positions). With all these efforts in mind, our aim here was to observe the AS process from a more global perspective, although based on quantitative parameters, $B_{sw}$, energy $E_{sw}$, and the activation energy $U$.

Before delving into the discussion, it is important to clarify that there are multiple trimeron excitations present. Each absorption of energy by magnetite corresponds to a reconfiguration of electronic states, i.e. some trimeron excitation. Among these excitations, the one which appears to be most intimately linked to AS is magnetocrystalline energy. When a magnetic field is applied, magnetic moments, or electronic orbitals tied to the spins via, mainly, spin-orbit coupling (SOC), are affected. Initially, the domain wall movement takes place, indicating that individual spins in trimerons incline towards the field direction (i.e. some elementary excitation takes place). Once completed, the volume magnetic moment partly rotates towards the field direction, causing the excitation of the trimerons from their normal state. It is important to note that such processes are not considered in our studies, although their relation to magnetic anisotropy is discussed further below in Sec. 4 d. Our focus solely revolves around crystallographic structure changes triggered by the magnetic field. It should be noted that there are additional structure-related excitations as mentioned in Ref. [32]. These include the appearance of *a*- and *b*-axis structural domains spontaneously appearing just below $T_V$, possibly related to additional *c*-axis twins (although not observed in Ref. [32]). Such considerations appear to be a natural extension of the phenomena we present here.

The main results of our study are presented in Figs. 2-4, with additional results for elevated pressure provided in SM [11].

Several noteworthy findings include the following:

**a. Temperature dependence of AS process**

When the temperature is increased, $B_{sw}$ and AS energy $E_{sw}$, which is closely related to $B_{sw}$ decrease. In each system, as $T$ increases, the occupation probability of higher-energy states also increases resulting in changes across the entire system. In the case a phase transformation, like the Verwey transition, is approached, not only do the energy levels become more populated, but this increased population also alters the states themselves, causing their energy to decrease. Consequently, their population continues to rise, ultimately leading to critical fluctuations and a phase transition. Each subsystem of the whole entity changes this way: it was shown, e.g., for trimeron elementary excitations [2] where the energy level separation of 5 meV decreased almost to zero just below $T_V$. The natural question arises: Is there any specific subsystem whose behavior dominates and acts as a leading force of the transition, shaping the temperature dependence of other subsystems?

Several phenomena can be considered here, such as the dependence of the spin-orbit coupling on $T$ [33] and lattice vibrations [35,36]. However, as the more detailed discussion in SM [11] elucidates, these factors appear to be unlikely contributors. Nevertheless, there is a 25% decrease in the electric field gradient with increasing $T$ for $C3$ and $C4$, representing two of the four components of the stoichiometric magnetite Mössbauer spectrum (see Fig. 3 in [34]). If the atoms constituting these components are actively involved in AS, as discussed further in Secs. 4 b and 4 d, and also in SM [11], our observation could potentially provide an explanation for the pronounced $T$ dependence of the AS.

The work required to switch the axis, $E_{sw}$, is strictly linked to $B_{sw}$ and is also $T$ dependent. Although it is a value associated with the collective process, i.e. an excitation of all trimerons simultaneously, the small value of the energy (0.25 K per Fe atom at lowest accessible temperatures down to 0.05 K at highest, close to $T_V$) is approximately 100 times lower than 5 meV reported in [2] also for some collective process engaging trimerons. This contrast implies that when employing a magnetic field, we are probing a distinct trimeron excitation compared to the one documented in [2]. In the latter case, the observed process involved the sliding of trimerons, whereas in our study, we observe their rotation as the dominant mode of excitation.

**b. Dopant and nonstoichiometry dependence**

Doping and nonstoichiometry, each accounting for less than 1/3% of the total iron content, considerably increase $B_{sw}$, AS energy $E_{sw}$, and $U$. Consequently, a greater effort is needed to reorganize the trimerons. This effect, vividly presented in Figs. 2-4, represents the most significant result of our studies and surpasses any potential parameter changes arising from the inevitable slight variations among the three nominally identical stoichiometric samples. This is equally strong, but opposite effect to $T_V$ vs. $x$ dependence, where also small amounts of defects considerably affect the transition, even changing its character (see Fig. 4 where the anticorrelation of $U$ with $T_V$ is presented. For the discussion of defects occupancy see Sec. 1 in SM [11]). This anticorrelation would suggest that "more rigid" trimerons make the Fe-B lattice more vulnerable to temperature disorder (lower $T_V$). This idea is not easy to accept bearing in mind that electrons within trimerons, and their confinement to the lattice (strong electron-phonon interactions) are usually reported in literature as being strictly connected with the mechanism of the Verwey transition.

However, it was shown in [37,38], the change of lattice distortion precedes the change of orbital and charge ordering on heating, as if electron ordering (charge in trimerons) sluggishly adjusted to already altered lattice. In case of further decrease of electron-lattice interactions, i.e. a possible effect of doping, electron charge may be changed by magnetic field with the lattice being inactive. This means that the easy axis does not change, i.e. there is no visible AS, despite some electronic rearrangement within trimerons. This idea is, however, not further supported by experiment because stoichiometric and slightly Zn-doped magnetite crystals show the same difference in lattice distortion and charge/orbital ordering temperatures, [37,38] yet their response to magnetic field, AS, is drastically different. Also, no clear change in electronic states is visible in slightly doped magnetite (as studied here) unlike in magnetite exhibiting continuous VT [39]. Therefore, we consider the above model as speculative, and the ultimate suggestion is that what we observe as the reorganization of electronic order (seen by us as a trimeron rearrangement) is not the same electronic reorganization as the one that takes place at $T_V$. This is additionally supported by the fact that $T_V$ lowers with the increase of the content of Zn, Ti, and $3\delta$ in another way than with the content of Al (see Fig. 1S(a) in SM [11]), while the results for $U$, $B_{sw}$ and $E_{sw}$ dependence on $x$ and $\delta$ are all very similar.

Doped magnetite was studied microscopically by NMR [40] and MS [34]. Although only a minor change of resonance field was observed in low Zn- and Ti-doped magnetite (with discontinuous VT) in NMR results, the rapid change of hyperfine magnetic field on Fe sites is visible once $x$ enters the second order Verwey transition regime, very close to the concentration we study here. This roughly coincides with the remark in [41] about highly selective oxidation of one particular $B$ position on Zn doping or nonstoichiometry, in about the same $x$ region where first order magnetite turns to second order. Note, however, that the effect of doping is element selective [40] even if the same Fe position is replaced, unlike in $T_V$ vs $x$ (for Zn, Ti, and $3\delta$) and as in our case of AS. However, the drop of the hyperfine field is universal.

The same conclusion of the drastic change of the system properties on doping when entering the second order regime is visible in MS results [34]: with increasing the temperature (below $T_V$), there is a substantial drop of the electric field gradient $V_{zz}$ in eight Fe octahedral positions (belonging to the $C$3 and $C$4 components mentioned above) with even a change of sign in one of those ($C$3). This is in accord with the increased value of activation energy $U$ in doped/nonstoichiometric samples that shows the pace of $B_{sw}$ changes with $T$ (the higher $U$, the more dependent on $T$ the switching field $B_{sw}$ is). The arguments for $C$3 and $C$4 involvement in AS are further shown in Sec. 4 d below.

No matter how subtle the electronic changes caused by doping/nonstoichiometry in seven measured samples might be, $B_{sw}$ behaves as if trimerons were pinned by defects caused by dopants/nonstoichiometry. Thus our results also show that any general claims drawn from nonideal crystals and concerning electronic trimeron structure can be questioned (see, e.g., [15]). The same phenomenon, of defects impact on material properties, is present in superconductors where defects pin vortices which makes current flow nondissipative.

The important role the defects (of several kinds) play in magnetite microstructure and the ensuing properties was already commented on, e.g., in [42], where the step in ac susceptibility was suggested to be due to magnetic domain walls pinning on crystallographic domains. Although those crystallographic defects are omnipresent in low-$T$ magnetite, the other defects may be additionally created by doping/nonstoichiometry. Our present results suggest that those kinds of defects pin trimerons in a more effective way than natural crystallographic domain walls do: $B_{sw}$ as well as $U$, are higher for defected samples.

Axis switching is a relaxation process; i.e., it depends on the time the field is applied to the sample. At sufficiently low $T$, the time of the axis change is long; in [13] the time dependence of magnetization was modeled and, using NMR, the authors observed the AS process which at 57 K required even 24 h to complete. Because of this time dependence, in order to provide similar experimental conditions to draw $B_{sw}$ from $m$ vs $B$, measuring time for each temperature should be adjusted accordingly. However, due to the time constraint this condition was not always fulfilled and the time of collecting the data was different for measurements at various temperatures. Nevertheless, the observation time does not challenge the fact that the trimeron system reacts to the external field much more readily in clean material than sluggishly as in doped samples. This fact justifies our estimation of all measured parameters in doped/nonstoichiometric samples; however the traces of this "time" problem are seen in our results. For example, this caused different $m$ vs $B$ slopes, visible, e.g., in Fig. 4S(d) in SM [11]. It also affected $B_{sw}$ defined as the first $B$ value where $m(B)$ was significantly off its value at $T = 4$ K. All these contributed to the uncertainty of $B_{sw}$ (and, subsequently, $B_{i\_sw}$) estimation.

Checking AS dependence on dopant concentration mixes two effects: the defects' impact on trimerons and the change of electronic structure these defects can cause. The effect of electronic structure on AS alone is probably better defined when hydrostatic pressure is exerted. This is seen in Figs. 2-4, 6, and 7 and discussed below.

**c. Pressure dependence of AS**

The literature on magnetite properties consistently suggests that the system is highly sensitive to small differences in structural parameters, including those induced by pressure. For instance, studies have shown that applying pressure can lead to an increase in the Néel temperature [43], indicating a strengthening of the octahedral-tetrahedral exchange coupling, which is an important parameter in understanding the possible magnetic origin of the Verwey transition [14]. The temperature dependence of elastic constants has also been investigated under pressure, revealing a slight decrease in the coupling of the order parameter to strain [44,45]. Moreover, pressure has been found to have a significant impact on the local electron ordering in the $e_g$ and $t_{2g}$ orbitals, even at temperatures close to room temperature [46]. Many studies have focused on the influence of pressure on the Verwey transition, and it has been consistently observed that $T_V$ decreases gradually with increasing pressure for pressure values below 1.5 GPa in stoichiometric magnetite, nonstoichiometric magnetite, and Zn-doped ferrites[2] [47,48,49,50] (see Fig. 13S(c) in SM [11] for some further results from the literature [51,52]).

Therefore, studying the influence of hydrostatic pressure on magnetite in the small range 0-1.2 GPa, could provide valuable insights into the peculiarities of AS and the subtle arrangement of orbital ordering.

It is important to note that applying pressure and subsequently releasing it does not leave the sample unchanged. The sample is different after a short stress which results, among other factors, in the increase of the $T_V$ under increasing impact strength, a still higher effect for more doped samples[3] [54]. Application of pressure alters the electronic structure of the crystal by reduction of atom-atom distances and when pressure is released, this effect disappears. However, pressure may also affect the defect state which may be irreversible. Since $T_V$ lowering is universally observed when the sample is measured under hydrostatic pressure, we conclude that the change of electronic states and the ensuing effect on VT prevails over the effect of defects. This was also the reason we used rather small pressure, where the effects mentioned above are minute.

The main findings of our research are presented in Figs. 2-4 (and in Fig. 12S in SM [11]). Additionally, pressure dependence of the switching field at representative temperatures (60 K for stoichiometric magnetite and 90 K for Zn-doped ferrite) is shown in Fig. 6(a). $B_{sw}$ seems to decrease with increasing hydrostatic pressure both in stoichiometric and Zn-doped magnetite. In other words, when pressure is applied, $B_{sw}$ drops in a manner similar to the decrease in $T_V$. A similar but less pronounced effect is observed in the AS energy $E_{sw}$ (Fig. 3).

Taking a closer look to Figs. 2, 3 and 6(a) (and 12S in SM [11]) it can however be noticed that after the initial drop of $B_{sw}$ and $E_{sw}$ from ambient pressure to $p = 0.37$ GPa, both parameters stabilize with $p$ with a further decrease for $p = 1.2$ GPa. Although it may be linked to some series of processes ocurring under pressure, the uncertainty of data points collected under pressure $p < 0.8$ GPa is rather high, and the unexpected phenomena take place at 1.2 GPa (see below). Therefore, the discussion of the subtleties in $B_{sw}$ and $E_{sw}$ vs $p$ is not justified and we only say that in general $B_{sw}$ seems to lower with pressure which is a general phenomenon in magnetite, independent of its doping character.

---

[2] Note, however, that $T_V$ rises with uniaxial pressure [53]. Also, in two papers, the initial rise, at $p < 0.2$ GPa was found for stoichiometric magnetite as also indicated in Fig. 13S(c) in SM [11].
[3] These studies were dedicated to assess meteorite impact on magnetite properties via VT observation.

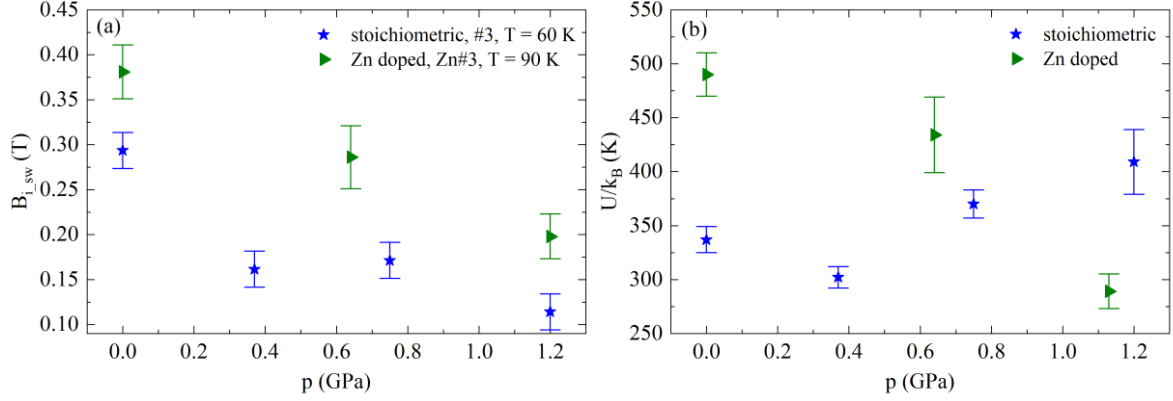

FIG. 6. (a) Pressure dependence of $B_{i\_sw}$ at two representative temperatures for stoichiometric (Fe$_3$O$_4$ #3) and Zn doped magnetite (Fe$_{3-x}$Zn$_x$O$_4$ Zn #3). The plots are cuts along the dotted lines in Fig. 2. Panel (b) presents the pressure dependence of $U$ measured in the same samples.

The factor that distinguishes stoichiometric and Zn-doped materials is the activation energy $U$ [Figs. 4 and 6(b)]. While pressure reduces $U$ in Zn-doped magnetite (resulting in a less temperature-dependent AS), the changes of $U$ for the stoichiometric sample are less evident, and there might even be a slight increase with pressure, although within a large range of error bars. Consequently, while $U$ shows a correlation with $T_V$ in Zn-doped magnetite ($T_V$ decreases with increasing $x_{Zn}$), some degree of anticorrelation is suggested for stoichiometric magnetite (including the region of very low pressure where $T_V$ appears to be at its maximum; see Fig. 13S(c) in SM [11]).

Hydrostatic pressure affects $B_{sw}$ and $E_{sw}$ in a similar manner to $T_V$: all decrease with pressure. This stands in stark contrast to the impact of doping/nonstoichiometry, which leads to a considerable increase in $B_{sw}$ while causing $T_V$ to decrease. Therefore, the working hypothesis emerges that the mechanism responsible for $T_V$ lowering with dopants/nonstoichiometry is distinct from the mechanism behind $T_V$ lowering with pressure, as already formulated above.

As already outlined in Sec. 3, the process of defining an easy axis through field cooling is not entirely effective in stoichiometric magnetite under pressures of 1.2 GPa. In this case, the magnetic easy direction appears to be better defined after applying the magnetic field along the initially unspecified direction [010]. Since the results are much more controversial than for other pressure conditions, more is presented in Sec. 5(b) (Fig. 10S) of SM [11], while the main result is discussed here.

Figure 7 displays the magnetization curves at $T = 100$ K along the "easy" direction [100] (after it was defined, as above, by field cooling with $B$ along [100]) and along the initially unspecified axis [010]. A comparison is made with $m(B)$ curves along the easy axis [100] and unspecified [010] at $T = 10$ K. It is evident that the two curves along [100] do not overlap, and the curve at $T = 100$ K exhibits a hysteretic behavior, resembling the behavior observed after the AS process. However, the repeated $m(B)$ scan along [100] is reproducible, indicating that no irreversible processes have occurred.

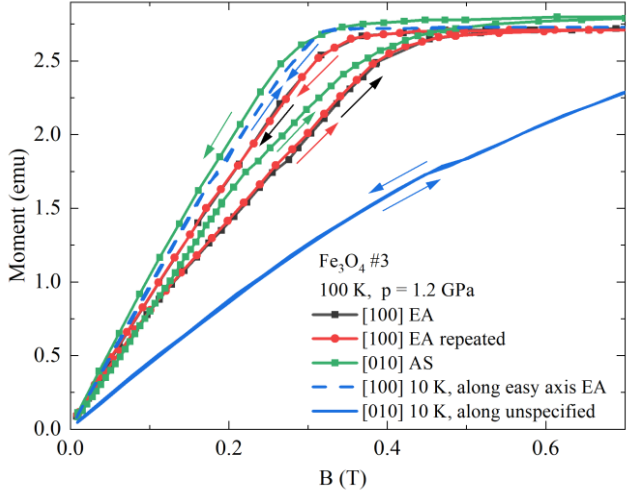

FIG. 7. $m(B)$ under $p = 1.2$ GPa at different temperatures. The data at 10 K prove that the $c$ and magnetic easy axes are well defined: dashed blue line is along an easy direction [100], while blue solid line is along an unspecified direction [010]. At 100 K the situation is different: $m(B)$ along the "easy" [100] shows a reversible hysteresis (black line), repeatable (red line). $m(B)$ along "unspecified" [010] (green) also shows hysteresis, but irreversible. Note that $m(B)$ on $B$ lowering is more steep along [010] (i.e. after "axis switching") than along the easy axis [100].

Interestingly, when the magnetization curve is measured along the initially unspecified direction [010], the AS process takes place, resulting in a final magnetization curve that resembles the "true" easy direction (depicted as the green curve in Fig. 7). The phenomenon initiates at even lower temperatures (see Fig. 10S in SM [11] where the results at 100 and 70 K are compared) where the hysteretic behavior of $m(B)$ along easy direction is less pronounced. A similar observation was made in a Zn-doped sample under pressure at 1.13 GPa and 105 K, as demonstrated in Fig. 11S(c) in SM [11], although this phenomenon was not extensively studied further.

In an attempt to comprehend the observed phenomena, several scenarios can be proposed. The prerequisite for the procedure used here to uniquely define the crystallographic $c$ and, simultaneously, magnetic easy axes was that electronic spins, manipulated by the external magnetic field, are so strictly confined to electronic states (i.e., also ionic arrangement) that the magnetic field uniquely defines the structure. This assumption, and the ensuing experimental procedure, were justified in ambient conditions but this may not be true at elevated pressure and high temperatures as our results show. A possible case is that either spins are less confined to the orbital moment (lower SOC), or orbital moment lowers with pressure, i.e., there is also magnetic field confinement to the ionic arrangement. Alternatively, magnetic axes no longer match crystallographic ones. In this last case, and assuming that the energy barrier between the "old" magnetic axes direction and "new" is not very high, the following explanation of the observed processes could be considered:

**First**, during the [100] field cooling, the structure still becomes uniquely defined, but magnetic anisotropy is smaller, resulting in a slightly different magnetic easy axis direction.

**Second**, magnetization vs $B$ along [100], although along the crystallographic $c$ axis, is along the axis that is not an easy magnetization axis and the magnetic domain structure is complicated and lacks saturation. However, for sufficient field (ca. 0.4-0.5 T) the saturation is achieved and the new energy minimum is realized. With field lowering, the magnetic system resides in this new energy minimum, even though the real one exists (but is separated by the energy barrier approximately proportional to the magnetic field) and is successively lowering: $m(B)$ is different than on field rising.

**Third**, in zero field the magnetic system switches to the "old" minimum and the repeated $m(B)$ is identical to the preceding one, as shown by the experiment (the red line in Fig. 7).

**Fourth**, when $m$ vs $B$ is measured in the [010] direction, the $m(B)$ curve is different than that along [100], eventually saturating in approximately the same field, but the saturation is realized by ionic changes, unlike along [100]. The argument for this is that the repetition of the same procedure gives different $m(B)$ curve, much more similar to the "ideal" $m(B)$ relation along the easy axis at ambient pressure. Note also that $m(B)$ on lowering is more ideal than that along the [100] direction.

The proposed explanation: complicated and strongly magnetic field dependent magnetic structure should be further tested, as well as the possible structural changes under this pressure at $T$ close to $T_V$.

**d. The relation to magnetic anisotropy**

The observed phenomena suggest that the magnetic anisotropy plays an important role in AS. In AS, electron orbitals in trimerons centers, which have more $Fe^{2+}$ character are altered, likely due to the direct interaction of the magnetic field with orbital magnetic moment or the interaction of the magnetic field with spin, later transferred to the electronic orbital by SOC. All these seem to be closely linked to the magnetic anisotropy energy ($E_a$). But magnetic anisotropy is only weakly temperature dependent [21,55,19] and marginally affected by small doping or nonstoichiometry (as seen in Fig. 15S in SM [11]), whereas AS exhibits significant changes under these conditions. Magnetic anisotropy involves placing the magnetic moments in a magnetic field which causes the reorganization of electronic states to lower their energy. This new energy under the magnetic field is higher anyway than without the field; therefore, finite energy is needed to rotate the magnetic moments. However, this process is reversible: once the magnetic field is removed, the electronic system returns to the initial state.

In the case of the processes studied here, if the energy rise of electronic states in the presence of magnetic field is higher than the barrier to an alternative electronic arrangement (in the form of trimerons), AS occurs. This AS process involves electron transfer leading to a new trimeron arrangement, accompanied by slight atomic displacements, ultimately resulting in a change in the $c$-axis orientation. In the defected magnetite, whether through doping or nonstoichiometry, the electron transfer leading to a new trimeron arrangement becomes more challenging. In other words, defects increase the energy barrier preventing trimeron rearrangement, while the rigidity of the orbital/electronic order (linked to the anisotropy energy $E_a$) remains largely constant. It is important to note that while magnetic anisotropy and axis switching are interconnected, they represent distinct phenomena.

The response to the external magnetic field was studied microscopically by observing individual Fe positions, including both the eight tetrahedral and 16 octahedral sites, using NMR [56], as well as groups of Fe positions using MS [12], all conducted under the influence of a magnetic field. The results demonstrated that the effective magnetic fields acting on Fe nuclei, both of isotropic and anisotropic character, change considerably with the direction of the magnetic field. Another aspect of these studies [56,57] highlights two groups of Fe $B$ positions ($B$1-$B$4 and $B$14 in one group and $B$7, $B$13, $B$16 in the other) that are highly sensitive to magnetic field compared to others. The calculated effective magnetic field in the first group [56] considerably decreases (by approximately 35%, compared to a maximum of 17% for other directions) when an external magnetic field is applied along <100>, different from the $c$ direction. Conversely, for the second group, the effective magnetic field considerably rises (again by 35%) under the same conditions. These two groups of atoms exhibit peculiar behaviors when observed by Mössbauer spectroscopy [34], all having relatively low valence (predominantly of $Fe^{+2}$ character). Among them, five positions ($C$3) are trimeron centers, with one position, $B$14, also serving as a trimeron end point [1,41]. In addition, all of these positions have a significant electric field gradient, much higher than other $B$ positions, the fact already mentioned in Secs. 4 a and 4 b above. These particular Fe positions are most sensitive to the effects of external field application along directions other than the $c$-axis, indicating their involvement in the AS process. Once AS starts, electron charge transfer to other $B$ sites of Fe occurs, leading to trimeron reorganization. This charge transfer was observed in resistance vs magnetic field studies as spikes in resistance [58,59].

**Conclusions**

In summary, the mechanism of switching both the magnetic and $c$ monoclinic axes (abbreviated here as axis switching, AS), which essentially involves the reorganization of trimeron order, was studied by the observation of magnetic moment vs magnetic field behavior in stoichiometric, nonstoichiometric, and Zn-, Al-, and Ti-doped magnetite single crystals, in some cases also under hydrostatic pressure $p < 1.2$ GPa. The level of doping and nonstoichiometry was kept below the highest values, $x$ and $3\delta \leq 0.012$, which ensured that the discontinuous character of the Verwey transition remained unchanged, thus avoiding the introduction of additional phenomena. The results were quantified by phenomenological parameters: the switching field $B_{sw}$, the activation energy $U$, and the energy needed to switch the axis $E_{sw}$.

Key findings from the studies include:

1. $B_{sw}$ exhibits a significant increase when the defects, in the form of either nonstoichiometry or dopants, are introduced into the magnetite lattice. In stoichiometric and Zn-doped crystals, an increase in hydrostatic pressure ($p$) leads to a decrease in $B_{sw}$.
2. A similar trend of increasing values with defects is observed for both the switching energy $E_{sw}$ and $U$. However, $U$ shows an anticorrelation with $T_V$, suggesting that AS and the decrease in $T_V$ vs. $x$ are not driven by the same microscopic mechanism. This is further supported by the fact that the work required to switch the axis is only on the order of 0.25 K per atom (and 0.05 K near $T_V$), rather than around 50 K as reported in [2]. This indicates that some other trimeron excitations are observed here, and since the excitations in [2] are associated with the Verwey transition, the ones we observe are not.
3. The application of pressure results in a lower value of $U$ for Zn-doped magnetite, whereas the effect of pressure on $U$ in stoichiometric magnetite is minor in comparison to the uncertainty of $U$. Similarly, the uncertainty in $E_{sw}$ does not allow for a firm estimation of the $E_{sw}(p)$ trend, however slightly, suggesting that it decreases with $p$.
4. Both $B_{sw}$ and $E_{sw}$ decrease with increasing temperature across all the samples, but the application of pressure mitigates this effect. The response of specific iron $B$ sites to temperature and magnetic field direction, such as sites $B$1-$B$4 and $B$14, and $B$7, $B$13, $B$16, as observed by NMR [56], or group of sites ($C$3 group and $C$4 group, respectively) as seen by Mössbauer spectroscopy [34], are certainly connected with the processes observed in this study and strongly indicate that the electronic states in these sites are mainly responsible for the AS process.
5. At pressures exceeding 0.8 GPa (particularly notable for $x = 0$) there is an observable change of the electronic energy vs magnetic field landscape, which is reflected in the reversible hysteresis in $m$ vs $B$ relation, with $B$ oriented along the "easy axis" predefined by the field cooling. When $m$ is measured with $B$ increasing along a magnetically unspecified direction, the evident irreversible and hysteretic changes, similar to AS, occur. However, the resultant $m(B)$ during decreasing $B$ is steeper than that for an easy axis. This suggests that the new direction becomes the "better" easy axis compared to the one enforced by field cooling.

**Acknowledgments**

The work was supported by the National Science Centre, Poland, Grant No. OPUS: UMO-2021/41/B/ST3/03454, the Polish National Agency for Academic Exchange under "Polish Returns 2019" Programme: PPN/PPO/2019/1/00014, and the subsidy of the Ministry of Science and Higher Education of Poland. M.A.G., K.P., I.B., Z.K., and W.T. acknowledge the support from the "Excellence Initiative – Research University" program for AGH University of Krakow. I.B. acknowledges support from the Swiss Confederation through the Government Excellence Scholarship.

Supplemental Materials for:

# The impact of hydrostatic pressure, nonstoichiometry, and doping on trimeron lattice excitations in magnetite during axis switching

T. Kołodziej[1,2], J. Piętosa[3], R. Puźniak[3], A. Wiśniewski[3], G. Król[1,4], Z. Kąkol[1], I. Biało[1,5], Z. Tarnawski[1], M. Ślęzak[1], K. Podgórska[1], J. Niewolski[1], M. A. Gala[1,6], A. Kozłowski[1], J. M. Honig[7], W. Tabiś[1,6*]

[1]AGH University of Krakow, Faculty of Physics and Applied Computer Science, Aleja Mickiewicza 30, 30-059 Kraków, Poland
[2]SOLARIS National Synchrotron Radiation Centre, Czerwone Maki 98, 30-392 Kraków, Poland
[3]Institute of Physics, Polish Academy of Sciences, Aleja Lotników 32/46, 02-668 Warszawa, Poland
[4]AGH University of Krakow, IT Solutions Centre, Aleja Mickiewicza 30, 30-059 Kraków, Poland
[5]Physik-Institut, Universität Zürich, Winterthurerstrasse 190, CH-8057 Zurich, Switzerland
[6]Institute of Solid State Physics, TU Wien, 1040 Vienna, Austria
[7]Department of Chemistry, Purdue University, West Lafayette, IN, USA
* Email: wtabis@agh.edu.pl

## 1. Samples characterization and the effect of doping

Our aim was to see how a small disturbance of magnetite crystal alter the trimeron excitations. Therefore, we slightly doped magnetite with nonmagnetic dopants or introduced minute number of vacancies, as shown below [1,20,3,4]:

Zn: $\equiv Fe_{3-x}Zn_xO_4 \equiv (Fe^{3+}_{1-x}Zn^{2+}_x)[Fe^{3+}_{1+x}Fe^{2+}_{1-x}]O_4$

Al: $\equiv Fe_{3-x}Al_xO_4 \equiv (Fe^{3+}_{1-0.15x}Al^{3+}_{0.15x})[Fe^{3+}_{1-0.85x}Fe^{2+}Al^{3+}_{0.85x}]O_4$

Ti: $\equiv Fe_{3-x}Ti_xO_4 \equiv (Fe^{3+})[Fe^{3+}_{1-2x}Fe^{2+}_{1+x}Ti^{4+}_x]O_4$

Nonst. $\equiv Fe_{3(1-\delta)}O_4 \equiv (Fe^{3+})[Fe^{3+}_{1+6\delta}Fe^{2+}_{1-9\delta}]O_4$

Here () denote tetrahedral (A) positions, while [] octahedral (B) ones. In the case of Al, this is the most probable alternative, although most probably $Al^{3+}$ enters both octahedral and tetrahedral sites [4]. The influence of these disturbing species on the lattice is very small, yet different. For example, while $Zn^{2+}$ (replacing Fe on A sites) and $Ti^{4+}$ (on B sites) cations expand the lattice, $Al^{3+}$ (entering both positions) shrinks it [5]. For lattice disturbance applied here this effect is, however, very small (ca. 0.01% for Zn and Al, possibly slightly larger for Ti), 10× lower than 1 GPa external pressure [22] used in case of our pressure studies. Contrary to this, the Verwey transition is a phenomenon vulnerable to even very small amount of spurious atoms, or nonstoichiometry as shown in Fig. 1Sa and documented in vast literature, see e.g. [7,8]. Already one third of a percent of Zn, Ti dopants or nonstoichiometry level changes the transition from discontinuous to continuous one and most studies claim that above 1% of doping the transition disappears completely. Note that this universal behavior is despite a fact that all those means to disturb the ideal structure enter this structure in a different way and with the different valence. The problem why a small amount of dopands/nonstoichiometry universally affect the transition is still not resolved although we have shown [9,10] that the excitation spectrum, observed with specific heat, is drastically different for two regimes of the Verwey transition. Note, finally, that in case of other dopants, e.g. Al used in this work, the effect of doping on VT is less acute (Fig. 1Sa) and does not obey the universal $T_V$ vs. $x = 3\delta$ relation.

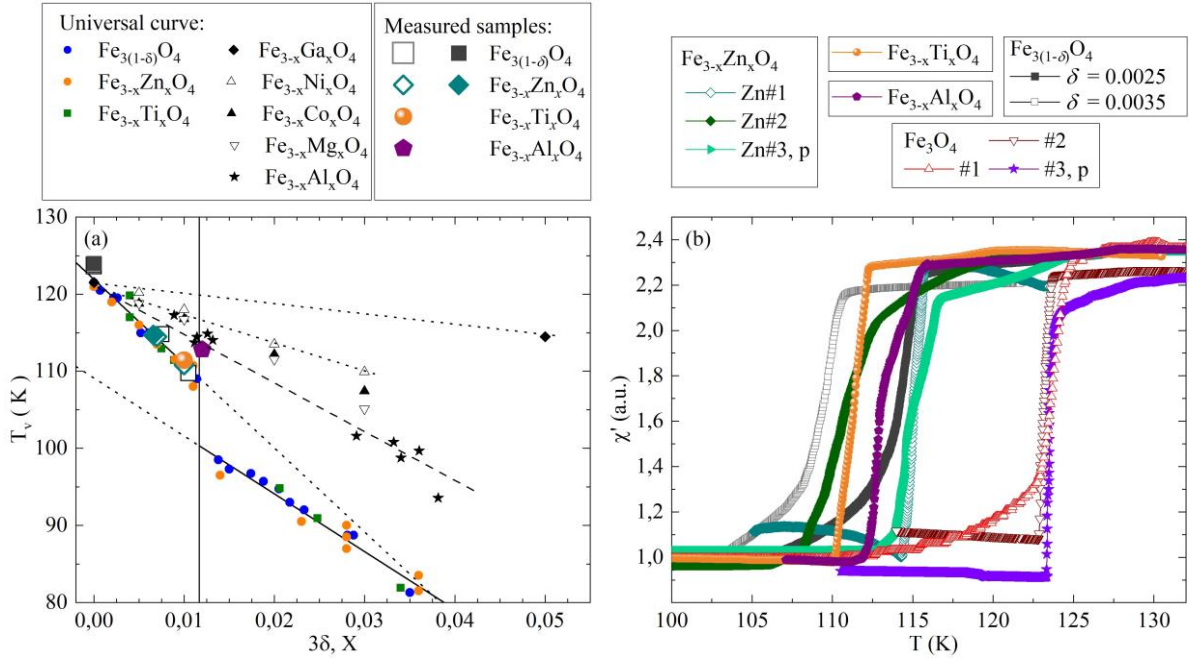

FIG. 1S. (a) Universal $T_V$ vs. $x$, $3\delta$ relation (for nonstoichiometric as well as Zn, Al, and Ti doped magnetite [1,20,3,4]). The data for Ga, Ni, Co, Mg doping are cited after [23]. Samples measured here are shown with large symbols (full in case measured under pressure). (b) Magnetic AC susceptibility vs. temperature for all measured samples.

The majority of the samples used in this study were cylindrical in shape, with a typical diameter of 2-3 mm and similar length, except for the measurements conducted under pressure, where the diameter was approximately 1 mm. The crystals were first oriented using Laue diffraction and subsequently polished with the long cylinder axis oriented along [001] cubic direction with the accuracy higher than 2 degrees. Since no further Laue check was performed after polishing the samples into cylinder, the direction mismatch of up to 3 degrees can be expected. Additionally, a spherical-shaped sample with diameter of 2.5 mm was oriented after its preparation.

The Verwey transition temperature of the samples was estimated from magnetic susceptibility $\chi_{AC}$ vs. $T$ relation, shown in Fig. 1Sb. Zn, Ti, and Al content was estimated from the universal $T_V$ vs. $x$ dependence presented in Fig. 1Sa, while the nonstoichiometry parameter $\delta$ in $Fe_{3(1-\delta)}O_4$ was set upon annealing.

## 2. Efficacy of field cooling in easy axis determination

The prerequisite in all experiments presented in the main text and SI is that $c$ crystallographic and, simultaneously, magnetic easy axes are uniquely defined by field cooling with $B$ along one of cubic <100> directions. Therefore, the check how large the cooling field should be was a first part of our studies. In case the axes are not uniquely defined magnetic moment vs. $B$, $m(B)$, along this axis is not typical for an easy axis direction and should show irreversible changes. In Fig. 2Sa, $m(B)$ are presented after field cooling in several fields showing that the sample should be cooled in at least 0.25 T field to set the easy axis.

However, field application (0.9 T) along one of <100> axes below $T_V$, e.g. at 80 K, also sets an easy axis in this direction (this is AS studied here), even though both processes (field cooling and forcing an easy axis at $T < T_V$) are different. In Fig. 2Sb, the comparison of the results (FC and 0.9 T field application after ZFC to 80 K) are presented. This shows that second process (bulk squares; it is supposed to define an easy axis along [100]) is almost as effective in easy axis determination as the field cooling in 0.25 T (triangles), or 0.25 T field cooling followed by field application along the same axis (open squares). Note, however, that slightly larger field is needed to complete AS in the first case (after ZFC and subsequent 0.9 T field application at $T < T_V$); this is presented in Fig. 2Sb. To assure that both ways to define easy (and $c$) axis lead to the same result, in all the studies presented in this paper, 0.9 T cooling field was used.

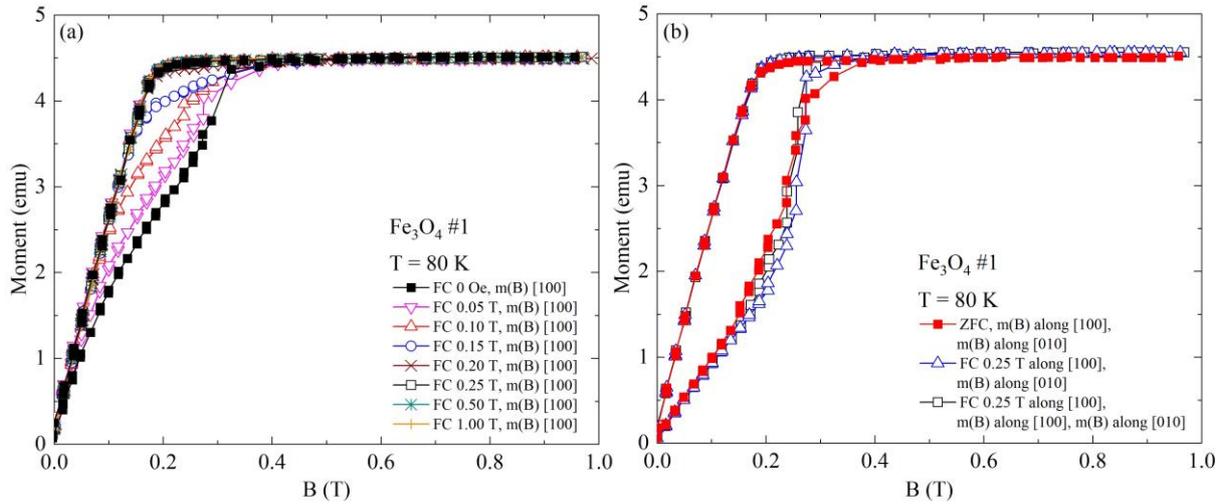

FIG. 2S. Efficacy of field cooling and field application below $T_V$ in establishing the $c$ and, simultaneously, easy magnetic axes. Field cooling in different fields along [100] defines, Fig. 2Sa, the easy axis (and $c$ crystallographic axis) above FC 0.25 T (starting from FC 0.2 T, $m$ vs. $B$ curves start to be identical to an easy direction case). Field cooling along [100] and 0.9 T field application along [100] at 80 K after ZFC have almost equal effect although slightly larger field is needed to complete AS in the first case (Fig. 2Sb).

## 3. Pressure experiments

The samples for pressure experiments (stoichiometric $Fe_3O_4$ #3 and Zn doped, $x_{Zn} = 0.0066$) were the long rods, (ca. 5 mm length, 0.8 mm diameter) cut with long rod axis along [100] direction. The samples were first loaded to the nonmagnetic (CuBe) cell filled with silicon oil (pressure transmitting medium) with a tiny piece of tin separated from the sample with nonmagnetic separating rod (Fig. 3S). Then, the cell was pressed to appropriate pressure (shown by the gauge), secured with the nut and attached to the measuring rod of VSM. The exact pressure was read from the superconducting tin transition ($T_C < 4$ K, i.e. below the typical temperatures the samples were measured). The details of the pressure cell are described elsewhere [12].

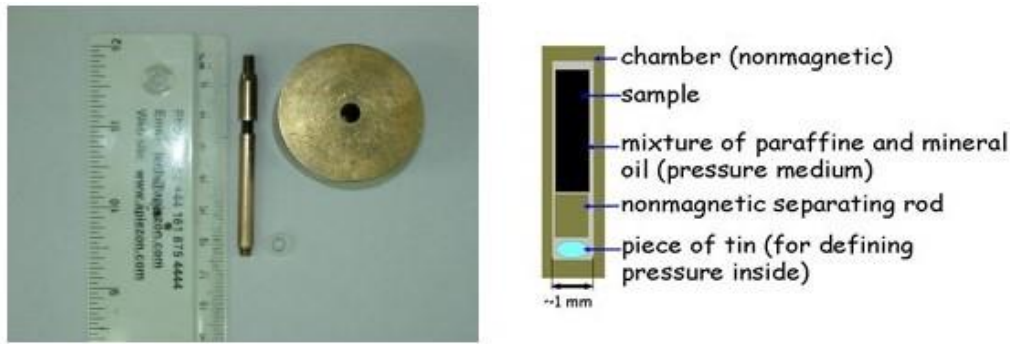

FIG. 3S. Pressure cell used in magnetization vs. pressure measurements. The pressure value was checked against pressure dependent superconducting transition in tin.

## 4. Results of *m* vs. *B* experiments under ambient pressure
### 4a. Stoichiometric and nonstoichiometric magnetite

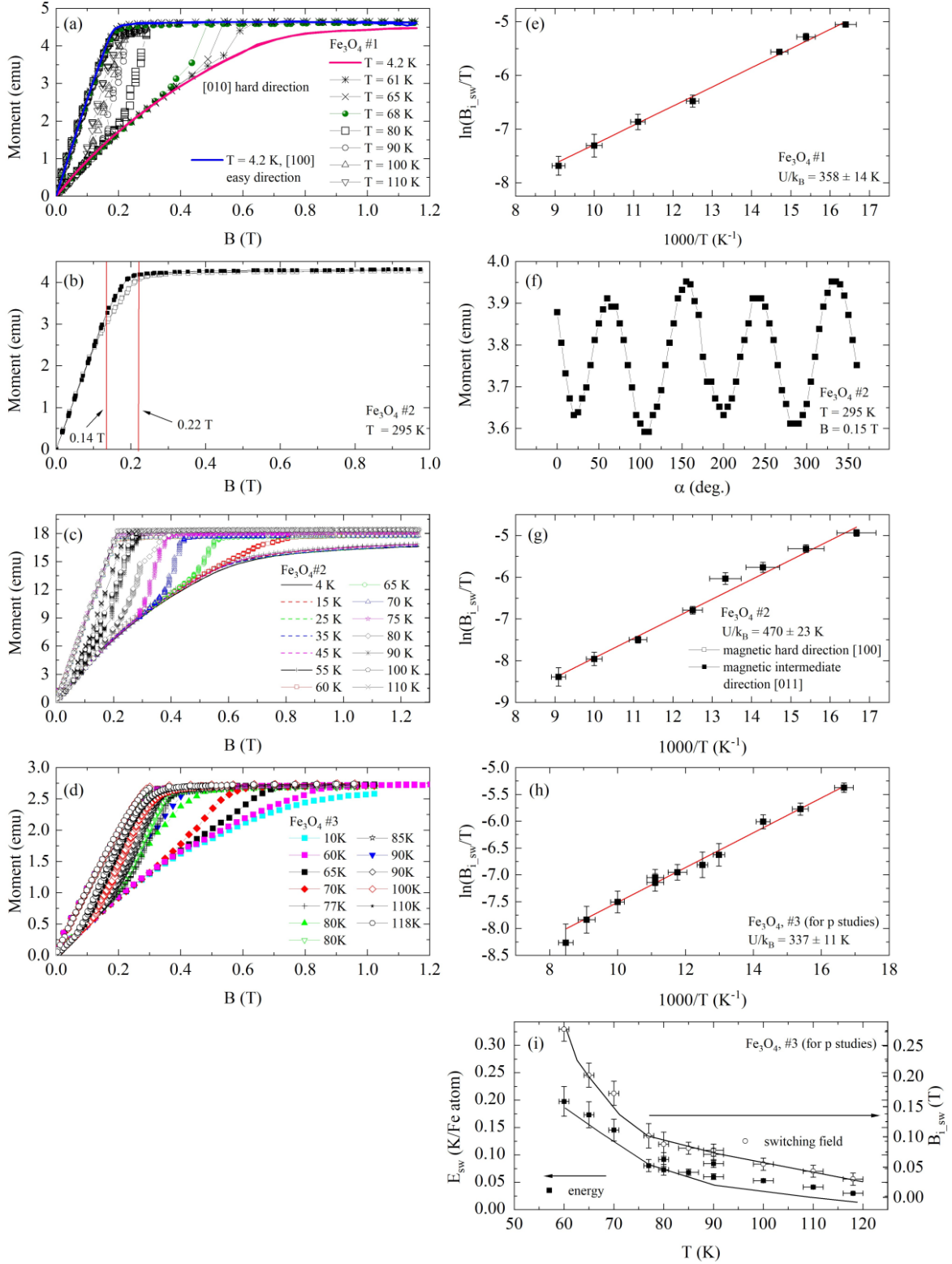

FIG. 4S. *m(B)* for stoichiometric magnetite below $T_V$ along unspecified direction showing AS (a, c, d), and ensuing fit to Calhoun's relation (e, g, h). In panels b and f *m(B)* for $Fe_3O_4$ #2 at 290 K, along <100>, hard above $T_V$, and <011>, intermediate above $T_V$ are presented. The difference between <100> and <011> are best seen in field of ca. 0.15-0.2 T. The minimum in *m* vs. vector *B* direction is along [100]. In case of $Fe_3O_4$ #3, also temperature dependence of the energy needed to switch the *c* axis, $E_{sw}$, as well as the temperature dependence of $B_{i\_sw}$ are shown (panel i).

## 4b. Nonstoichiometric $Fe_{3(1-\delta)}O_4$ magnetite

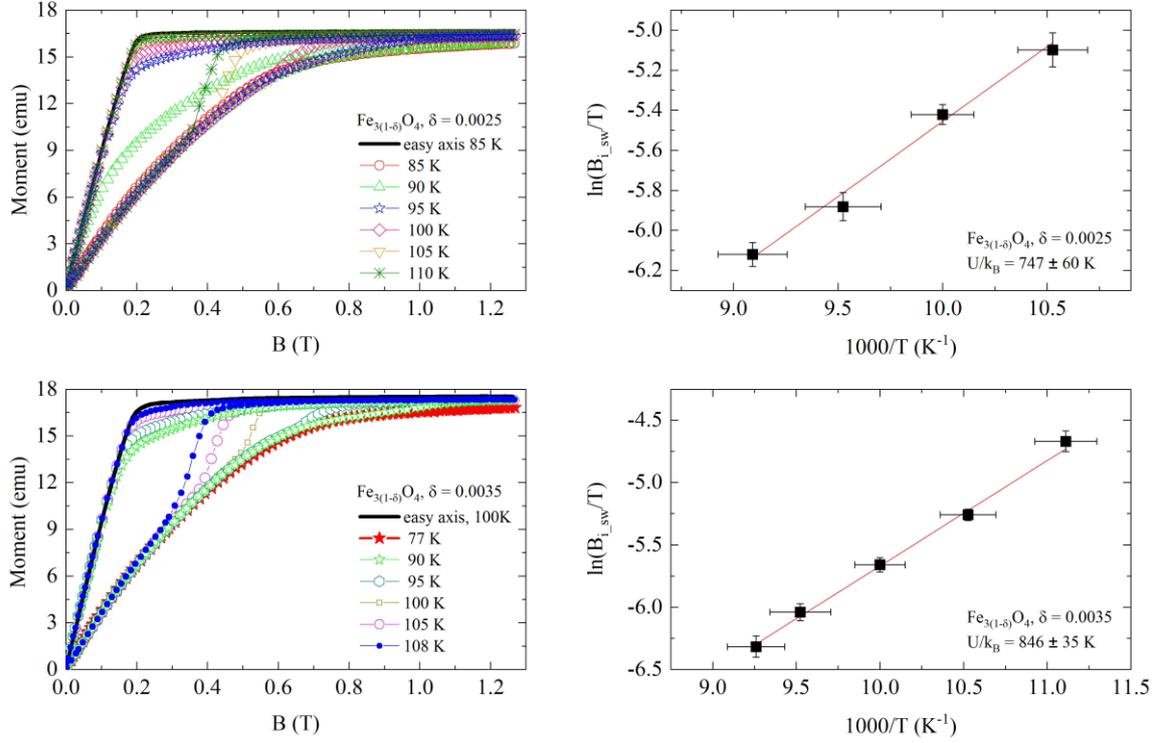

FIG. 5S. Same as in Fig. 4S but for nonstoichiometric samples. All the curves, except the indicated one, are along unspecified axis and show AS.

## 4c. Zn, Al, and Ti doped magnetite

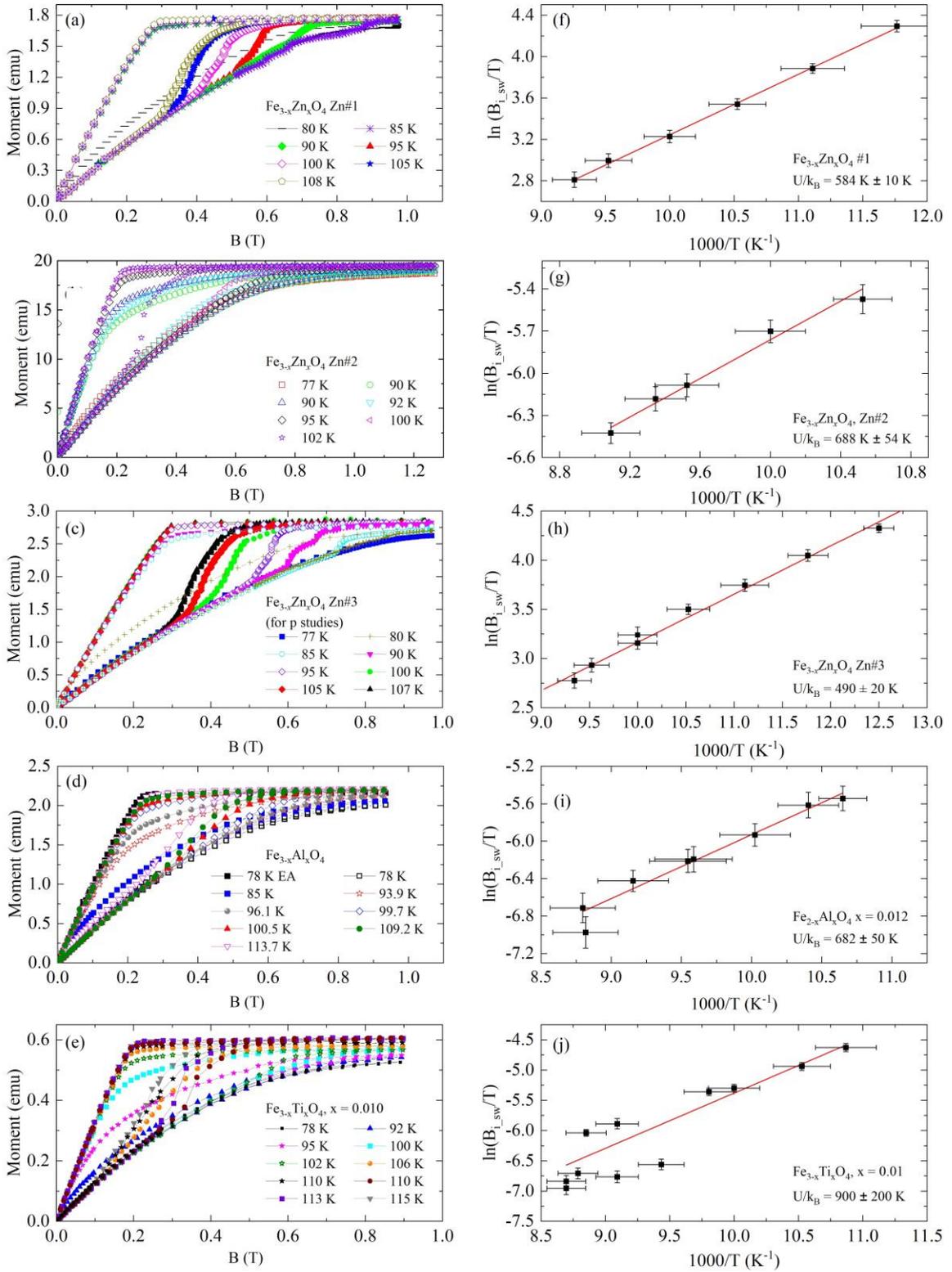

FIG. 6S. $m(B)$ (panels a-e) and fits to Calhoun's relation (f-j) for doped samples in ambient pressure. Since the data for Ti doped sample are of lower quality, in this case also those $m(B)$ curves where $B_{sw}$ estimation was not possible are presented. Note also that the uncertainty of $U$ is, in this case, large, unlike in other samples.

In Fig. 7S the *T* dependence of switching field and energy needed to switch the axis for all measured samples under ambient pressure are shown.

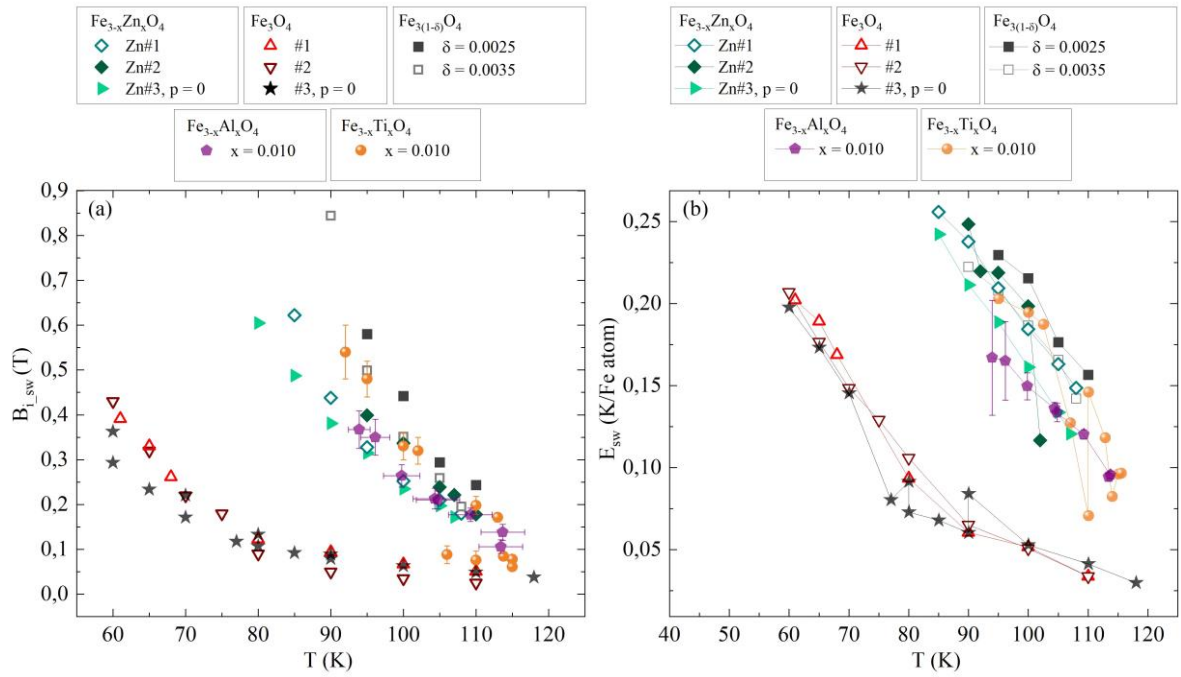

FIG. 7S. *T* dependence of switching field (a) and energy needed to switch the axis (b) of all measured samples under ambient pressure.

## 4d. Sample training experiment

Axis switching is an activation-type phenomenon, as is clear both from old [13] and our [14,15] studies. It is thus important to check how repeatable this process is and how it depends on the measurement procedure. We have performed AS five times, after the single field cooling along [100] direction, with the field first directed along [010], then [100], etc. Each time AS occurred, magnetic field direction became a new the $c$ axis that could be again switched by repeated field application. The results of this sample training, for a representative temperature (80 K and 100 K), fitting to the Calhoun's formula and $B_{i\_sw}$ vs. $T$ for three successive runs are shown in Fig. 8S. This is clear that switching fields are highest for the first switching event and are lower, and almost constant, in subsequent switching, while temperature dependence of it (expressed by the activation energy $U$) stays constant ($U/k_B$ = 364 K, 357 K, 329 K, all within the error bar of ca 50 K; this is presented in Fig. 8Sc).

The results from these experiments, i.e. the decrease of switching field with no considerable changes of $U$, show that these two parameters are not straightforwardly coupled as Calhoun's formula suggests. In other words, defects caused by AS decrease $B_{sw}$ without affecting the barrier the charges (and ions in a slight ionic rearrangement) must cross. $U$ also measures how the process reacts to the changes of temperature; the result suggests that smaller field is needed to trigger AS, but the pace it changes with $T$ and the barrier are the same. The most important outcome of this experiment is that AS introduces some changes to the system ("AS defects") that act unlike dopants/nonstoichiometry do.

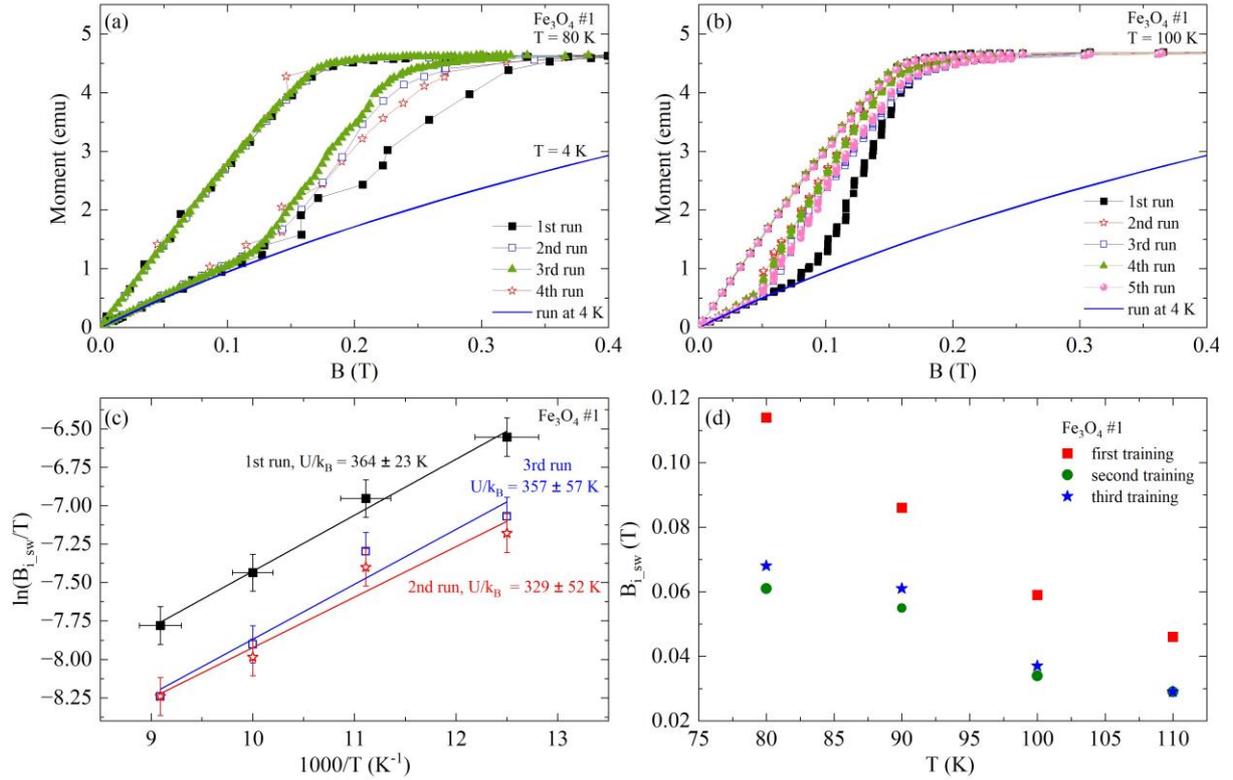

Fig. 8S. Sample training. Panels a and b: selected data of $m$ vs. $B$ from which $B_{i\_sw}$ vs. $T$ and $U$ were calculated. In each case, the sample was first FC in 0.5 T along [100] and then $m(B)$ was measured along [010] setting this as an easy axis. The sample was then rotated 90° (to an unspecified axis) and the procedure $m(B)$ was repeated ("2 run"). The procedure was repeated 4 (panel a) or 5 times (panel b). Fitting to Calhoun's formula is presented in panel c, while switching field vs. temperature during the sample training is in panel d.

## 5. Results of pressure experiments

### 5a. *m*(*B*) experiments for stoichiometric $Fe_3O_4$ #3

Since samples had slightly lower diameters than the sample chamber (this is shown in the scheme of the pressure cell in Fig. 3S) some movement in silicon oil was possible and was observed. Therefore, as already remarked in the main text, the procedure to find cubic hard (easy below $T_V$) and intermediate directions was done at 160 K to allow pressure transmitting medium to freeze. Also, the size incompatibility could result in some skew of the sample axis and, consequently, in some difference between magnetic moment in two perpendicular directions. This problem, although noticed in some cases, is not further discussed.

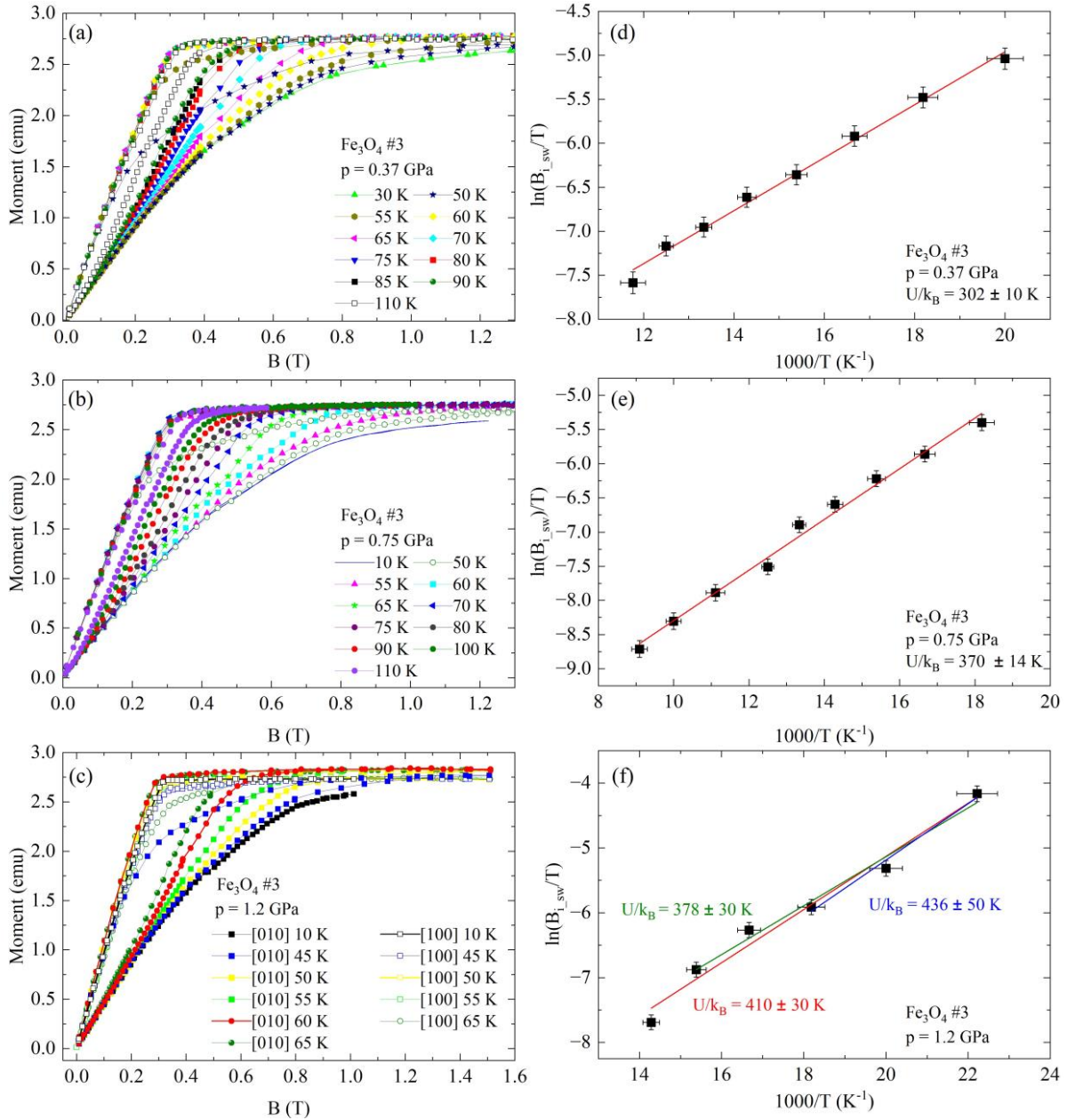

FIG. 9S. The results of pressure (a, b, c) measurements of the stoichiometric sample $Fe_3O_4$ #3 (the data under ambient pressure are presented in Figs. 4Sd and 4Sh). In panels d, e, f $\ln(B_{i\_sw}/T)$ vs. 1000/*T* and ensuing *U* values are shown. Since for *p* = 1.2 GPa the clear scheme of [100] as an easy axis and [010] as an unspecified one starts to be uncertain, *m* vs. *B* for both directions are presented.

## 5b. Monoclinic *c* and magnetic easy axes cannot be simultaneously set for Fe$_3$O$_4$ #3

As mentioned in the main part, the procedure to set an easy direction along cooling field was effective in case $p$ = 1.2 GPa and for $x$ = 0 at lower $T$ below 65 K. However, at higher temperatures, above 70 K, magnetic moment $m(B)$ along [100] (supposed to become an easy axis after FC) did not show typical results as for an easy axis.

This is presented in Fig. 10Sa at 70 K and compared to the usual behavior (an unspecified [010] at 10 K and an easy direction [100] at 30 K under ambient pressure). There is a hysteresis at 70 K in $m(B)$ along "easy" [100] (but reversible, as discussed in the main text), but $m(B)$ along [010] shows the process similar to AS with $m(B)$ under lowering $B$ steeper than an "easy" axis. The repeated $m(B)$ along the same direction (it should be on the same line if AS were completed) is slightly different as if the easy axis was "strengthened".

The effect is more visible at higher temperatures what Fig. 10Sb shows. In particular $m(B)$ at 100 K on field rising for an unspecified [010] is steeper than for "easy" [100] direction.

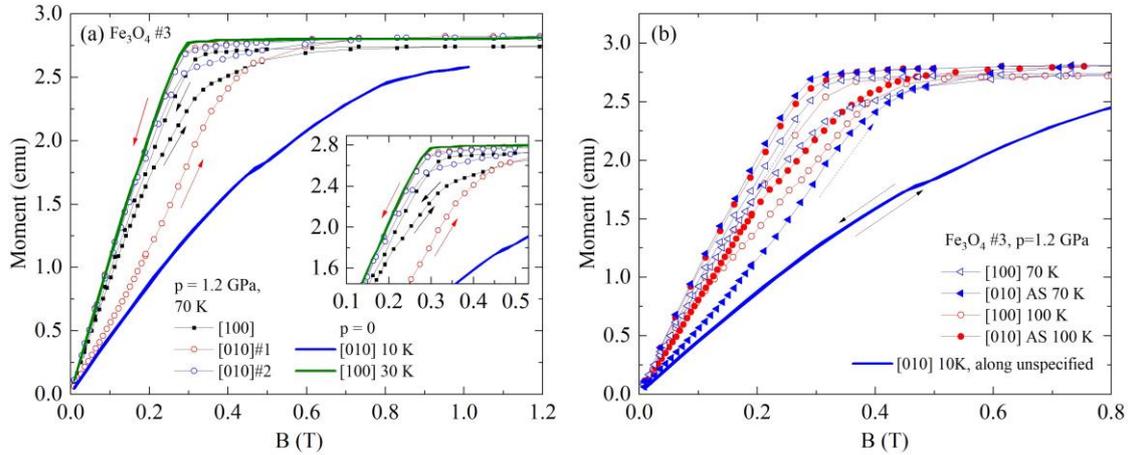

FIG. 10S. Magnetization process of stoichiometric sample Fe$_3$O$_4$ #3 under pressure of 1.2 GPa and at $T$ > 65 K showing different than in other cases behavior. First, panel a, $m(B)$, along the "easy" [100] shows hysteresis, increasing at higher temperatures (panel b) and reversible (Fig. 7 in the main text) when repeated. Second, $m(B)$ along "unspecified" [010] also shows hysteresis that is non reversible (panel a: red and blue symbols). Note that $m(B)$ on $B$ lowering is steeper along [010] (i.e. after "axis switching") than along the "easy axis" [100]. The process increases at higher temperatures as panel b and Fig. 7 in the main text show.

## 5c. *m*(*B*) experiments for Fe$_{3-x}$Zn$_x$O$_4$ Zn#3

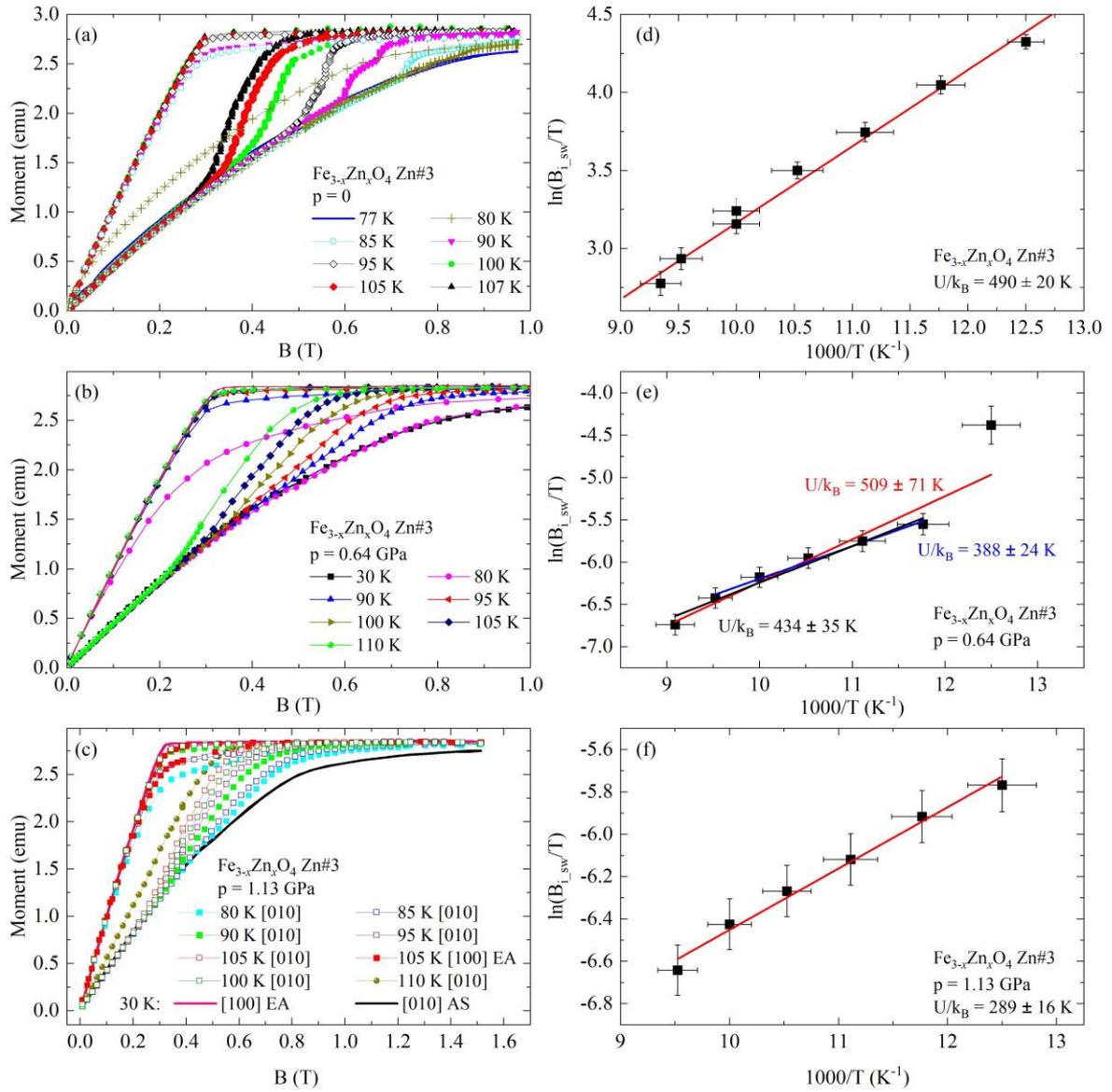

FIG. 11S. The results of ambient (a, d) and pressure (b, c, e, f) measurements of the Zn doped sample and fittings to Calhoun's relation. Note, panel c, that ill-defined easy axis starts to be visible also in Zn doped sample under pressure 1.13 GPa at 105 K.

## 5d. Pressure impact on energies and $B_{sw}$

In Figs. 12Sa, b the same parameters as in the main text are presented but focused only on dependence on pressure.

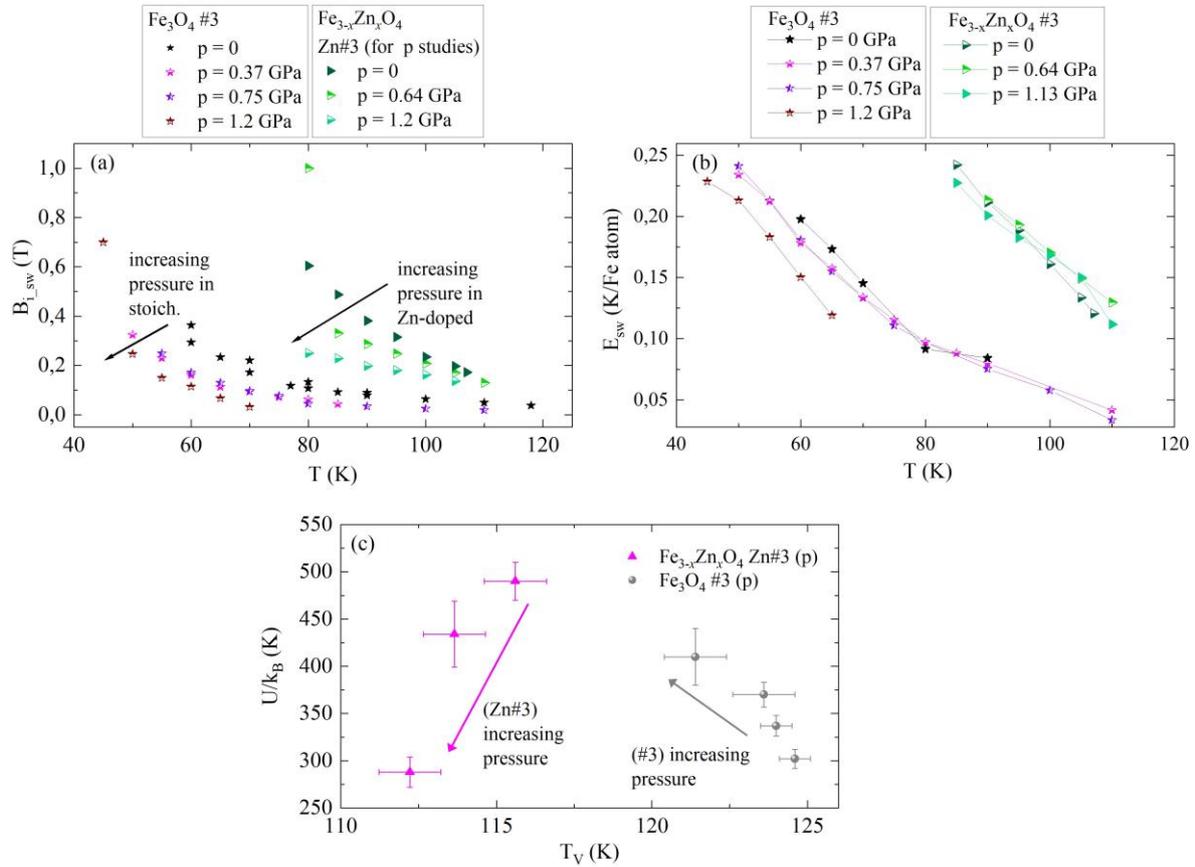

FIG. 12S. Temperature dependence of the switching field (a), energy needed to switch the axis (b) and the correlation of activation energy with the Verwey temperature (c) for both samples measured under elevated pressure.

## 5e. $T_V$ dependence on pressure

The measurements were performed in non-saturation field and were solely aimed on $T_V$ vs. $p$ check. Note that the $T_V$ for $p = 0$ was not recorded but was drawn from susceptibility measurements. In any case, our result confirms the result from [52] of a possible initial rise in $T_V$ vs. $p$.

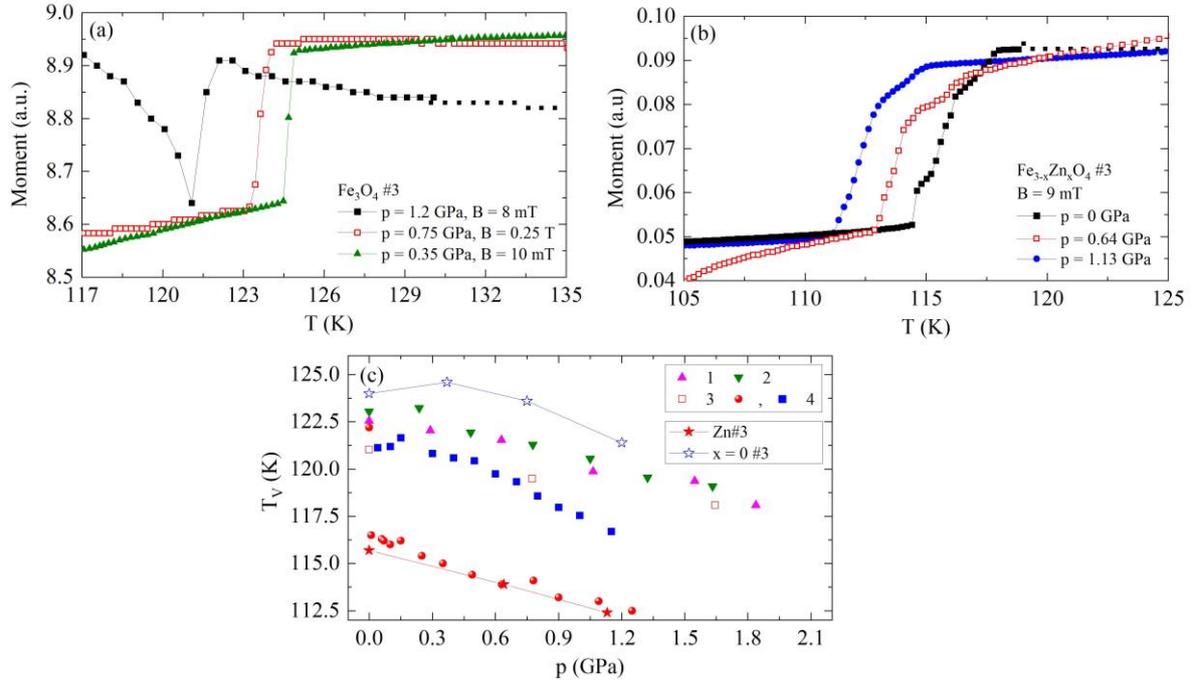

FIG. 13S. $T_V$ vs. pressure for measured samples and the literature data. In present studies $T_V$ was drawn from magnetic moment vs. $T$ dependence on heating (panels a and b). In panel c those $T_V$ values (stars) are superimposed on the literature data of low pressure impact on Verwey transition temperature (1: [16] 2: [51], 3: [18], 4: [52]).

## 5f. Pressure dependence of magnetic moment

The important information presented in the literature few years ago concerned the change of spinel structure under pressure: it was suggested in [25,26,27,28] that apart from the Verwey transition there is also a transition from high $T$ inverse spinel to the low $T$ normal spinel lattice $[Fe^{2+}]_{TET}[Fe^{3+}Fe^{3+}]_{OCT}$ that proceeds at the same temperature $T_{CC} = T_V$ as the Verwey transition under ambient pressure, but with $T_{CC}$ growing when pressure is increased. Such coordination crossover should result in a drastic (20%) change of saturation magnetic moment vs. $T$ relation. Although some contortions in $M_s$ vs. $T$ are present, 5%, at most, (Fig. 14S) the pronounced changes were not found in our results. Thus, in agreement with other literature [29,30], our data question the exact character of inverse to normal spinel transition. However, since we see an unexpected behavior of the magnetic easy axis, as commented in the main text, some processes under pressure certainly take place possibly related to the recently mentioned changeover in cation presence in A and B positions [31].

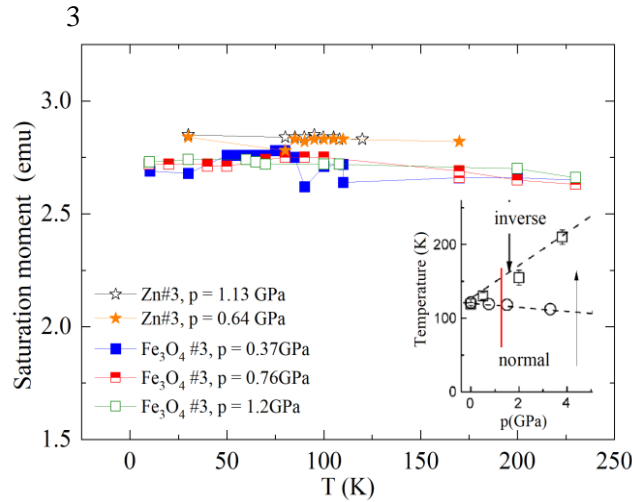
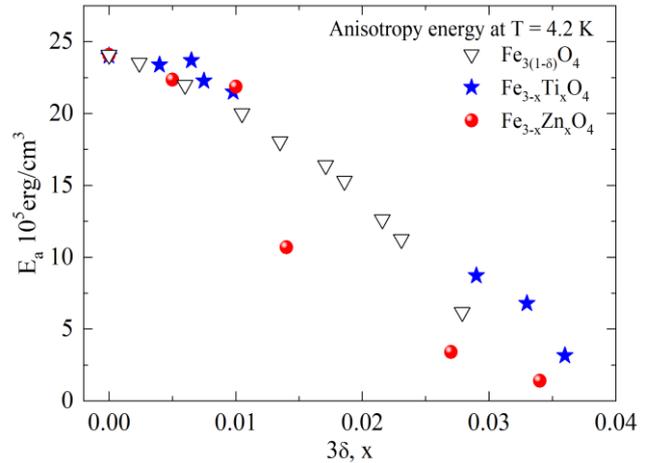

FIG. 14S. Saturation moment $m_S$ vs. $T$ for studied samples. No major change of $m_S$ was found that refutes claims of the pressure-induced change from low-$T$ normal to high-$T$ inverse spinel at $T_{CC}$. (shown in the inset; here red line limits the $p$ range covered here).

FIG. 15S. Magnetic anisotropy energy for stoichiometric, nonstoichiometric, Ti doped and Zn doped magnetite (based on [1,3,27]).

## 6. Relation to magnetic anisotropy

Axis switching is linked with magnetic anisotropy. Therefore, the temperature dependence of magnetic anisotropy drawn from published literature data is presented in Fig. 15S.

## 7. Additional information on the temperature dependence of AS

Several phenomena can be considered that trigger temperature dependence of AS.

First, it was suggested that spin-orbit coupling may depend on $T$ [33]. In the case of such situation, the spins in octahedral positions might gradually increase their coupling to electronic orbitals. This coupling may culminate in some magnetic field and at some temperature in a coupled spins and orbitals reconfiguration, i.e. AS. However, this should presumably be linked to some increase of saturation magnetic moment in magnetite below $T_V$ (that would result from orbital moment alignment along magnetic field), the phenomenon never observed. We therefore consider this mechanism of AS temperature dependence as rather improbable.

The next species that certainly change with temperature are lattice vibrations. Although phonon DOS does not seem to be $T$-dependent and even the change of it at $T_V$ is not very pronounced [10,35], it is natural to suspect that increasing the amplitude of atomic vibrations should increase the probability of AS and also lower $B_{sw}$. Since vibration amplitude has roughly square root dependence in $T$, the same dependence should characterize $B_{sw}$ vs. $T$. However, $B_{sw}$ vs. $T$ has an activation, exponential, character already used by us for activation energy $U$ estimation. Thus, phonons alone cannot be mainly responsible for the phenomenon we observe.

Finally, AS and its change with temperature can be related to 8 octahedral Fe positions, B4, B14, B3, B2 and B1, and B7, B13 and B16 (see [34] for the reference to Fe sites labeling), which are grouped into components C3 and C4, respectively, in the Mössbauer spectrum of stoichiometric and Zn-doped magnetite [34]. Atoms in these two groups experience a particularly large decrease in the electric field gradient $V_{zz}$ with $T$ below $T_V$. There is also a large change in $V_{zz}$ for these sites when Zn doping exceeds the first order transition limit, with the sign of $V_{zz}$ even reversed for the C3 group and a simultaneous increase in temperature dependence. Of these positions, one group, C4, has lower $B_{eff}$ (36 T compared to about 50 T for other groups) and both groups have the lowest valence, close to 2+ (the highest isomer shift parameter). Also, both groups are peculiar in the sense that their anisotropic component of $B_{eff}$ is relatively high and the isotropic component low, in contrast to other positions, and that a sharp decrease of $B_{eff}$ with $T$ is seen for C4 at higher Zn contents. All this suggests that a strong $T$ dependence of AS may be somehow related to these B positions.